\pdfoutput=1
\documentclass[10pt,compsoc]{IEEEtran}
\usepackage[utf8]{inputenc}
\usepackage{fp}

\usepackage{balance}

\usepackage[numbers]{natbib}
\usepackage{multirow}
\usepackage{hyperref}
\usepackage{color,soul}

\ifCLASSINFOpdf
  \usepackage[pdftex]{graphicx}
  \graphicspath{{images/}}
  \DeclareGraphicsExtensions{.pdf}
\else
\fi
\usepackage{amsthm}
\usepackage{amssymb}
\usepackage{bm}
\usepackage{amsmath}
\usepackage{svg}

\usepackage[linesnumbered,ruled]{algorithm2e}
\usepackage{listings}
  \usepackage[caption=false,font=footnotesize]{subfig}

\newlength{\codeindent}
\begin{document}
%
\title{Coherence Traffic in Manycore Processors with Opaque Distributed Directories}
%
%
%
%

\author{Steve~Kommrusch,
		Marcos~Horro,
		Louis-No{\"e}l~Pouchet,
		Gabriel~Rodríguez,
		Juan~Touriño,~\IEEEmembership{Senior Member,~IEEE}
\IEEEcompsocitemizethanks{\IEEEcompsocthanksitem S. Kommrusch and L.-N. Pouchet, Deparment of Computer Science, Colorado State University, USA.\protect\\}
	\IEEEcompsocitemizethanks{\IEEEcompsocthanksitem M. Horro, G. Rodríguez (corresponding author), and J. Touriño, Universidade da Coruña, CITIC, Computer Architecture Group; Campus de Elviña, s/n, 15071 A Coruña, Spain; email: gabriel.rodriguez@udc.es.\protect\\}}

\IEEEtitleabstractindextext{%
\begin{abstract}
  Manycore processors feature a high number of general-purpose cores designed to work in a multithreaded fashion. 
  Recent manycore processors are kept coherent using scalable distributed directories. A paramount example is the Intel Mesh interconnect, which consists of a network-on-chip interconnecting ``tiles'', each of which contains computation cores, local caches, and coherence masters. The distributed coherence subsystem must be queried for every out-of-tile access, imposing an overhead on memory latency. This paper studies the physical layout of an Intel Knights Landing processor, with a particular focus on the coherence subsystem, and uncovers the pseudo-random mapping function of physical memory blocks across the pieces of the distributed directory. Leveraging this knowledge, candidate optimizations to improve memory latency through the minimization of coherence traffic are studied. Although these optimizations do improve memory throughput, ultimately this does not translate into performance gains due to inherent overheads stemming from the computational complexity of the mapping functions.
\end{abstract}

}

\large{This work has been submitted to the IEEE for possible publication. Copyright may be transferred without notice, after which this version may no longer be accessible.}

\maketitle

\IEEEdisplaynontitleabstractindextext

%
\IEEEpeerreviewmaketitle

\IEEEraisesectionheading{\section{Introduction}\label{sec:introduction}}


%
%
%
%

\IEEEPARstart{M}ANYCORE processors feature a high number of general-purpose cores designed to work in a multithreaded fashion. In order to make systems scalable, current designs are usually based on replicated IP blocks connected by a high-performance fabric. An example of such an approach is the Intel Mesh interconnect (IM), first featured in the Intel Xeon Phi Knights Landing (KNL) processor~\cite{Sodani:HotChips2015}. The IM is the current interconnection standard in the most advanced Intel processors, including Intel Xeon Scalable servers and the High-End Desktop family of Core-X chips~\cite{Arafa:CascadeLake,Tam:SkyLake}.

\begin{figure}
	\centering
	\includegraphics[width=.8\columnwidth]{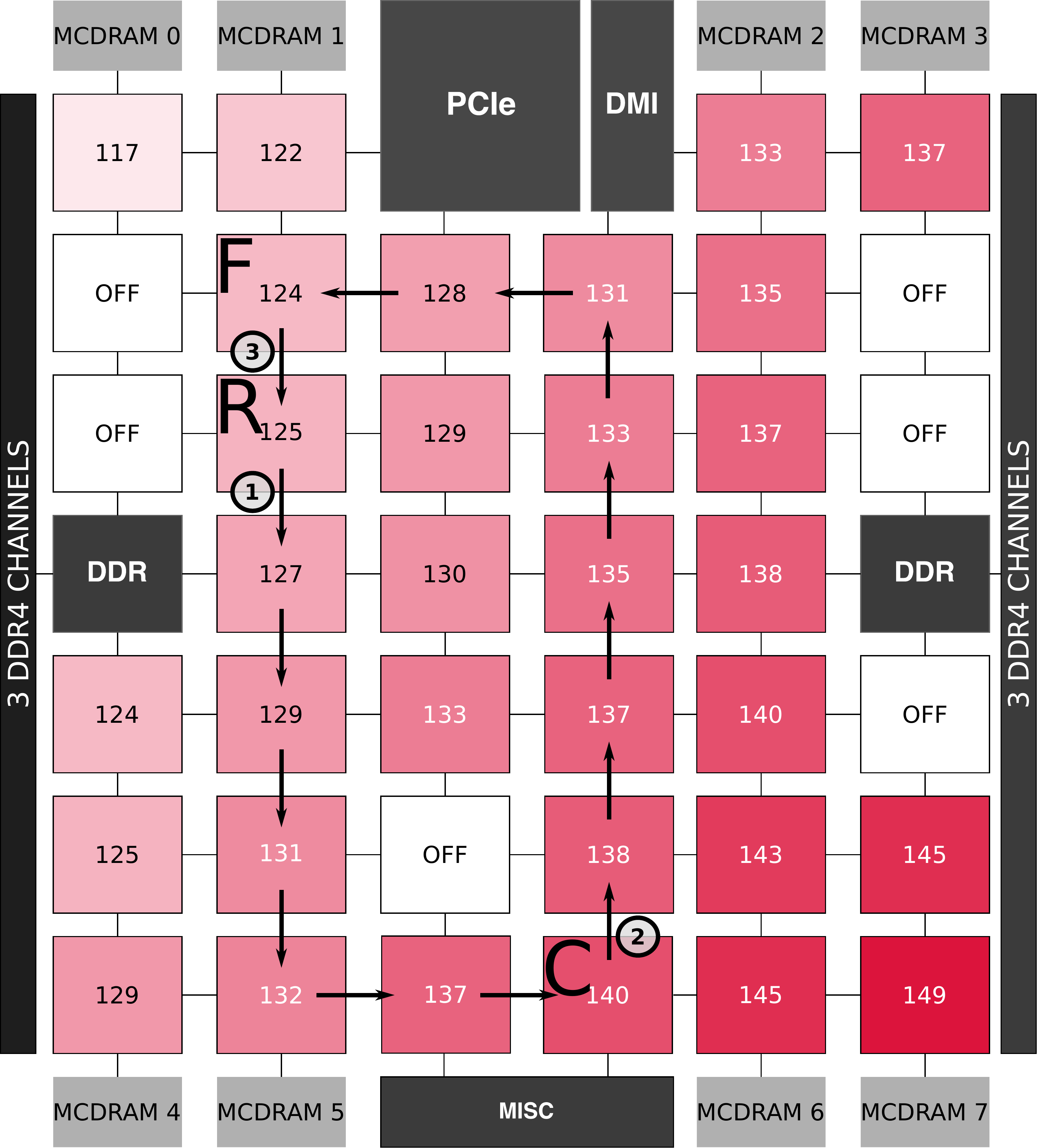}
	\caption{Actual access latencies from each tile in the mesh of an Intel x200 7210 processor to MCDRAM \#0, for a memory block whose coherence data is contained in the tile next to the memory interface. We note differences in access latency of up to 32 CPU cycles (a 27\% overhead over the minimum observed latency of 117 cycles).}
	\label{fig:Intro:KNL-Latencies}
\end{figure}

Each IP block in an IM-based processor, called ``tile'', includes computation cores and local caches. In order to maintain memory coherence the system employs the Intel MESIF protocol, supported by a distributed directory. Each tile includes part of this distributed directory in a component called the Caching/Home Agent (CHA). The directory must be accessed each time a core requests access to a memory block which is not already locally available in the appropriate state. This distributed design increases the scalability of the coherence system by removing the bottleneck that a centralized directory would impose, but causes a non-uniform increase in the network latency due to the varying distances between a tile and the set of CHAs on the mesh. Figure~\ref{fig:Intro:KNL-Latencies} details this latency variation across the full mesh of an Intel Xeon Phi x200 7210 (Knights Landing). As can be observed, latency overheads are higher than 25\% for extreme cases.

One key aspect of the design of the CHA-based directory in IM processors is that both the physical layout of the logical components of the processor and the mapping of memory blocks to CHAs are opaque and non-disclosed by Intel. This prevents memory latency optimizations, since the programmer has no a-priori knowledge of the latency that can be expected for each access. Furthermore, Intel advertises this architecture as UMA, since the \emph{average} memory access latency is approximately the same for all tiles in the mesh. This article builds on a previous work~\cite{Marcos:DAC19}, which reverse-engineered the physical layout of the logical components of the processor and showed how this knowledge, coupled with an inspector-executor which dynamically analyzes which CHAs are associated to each memory block access, can be used to optimize irregular codes. Leveraging this, the present work focuses on building a closed form function of the mapping of memory blocks to CHAs in order to remove the costly inspection phase, enabling new optimization strategies for IM processors. More specifically, this paper makes the following contributions:

\begin{itemize}
	\item The mapping of memory blocks to CHAs is reverse-engineered. Binary functions which compute a target CHA from a physical memory address are exposed and shown to be pseudo-random in nature (Section~\ref{sec:ReverseEngineering}).
	
	\item Different optimization strategies to improve memory latency by leveraging the mappings between memory and CHAs are designed. Approaches are proposed based on both dynamic and static work scheduling (Sections~\ref{sec:RuntimeOptimization} and~\ref{sec:CompileTime}).
	
	\item Experiments are performed to quantify the effectiveness of the proposed optimizations. It is shown how the proposed schedulings improve the memory latency by exploiting CHA proximity. However, due to the pseudo-random nature of the block-mapping functions the implementation of these schedulings affects other performance-impacting factors, which may ultimately lead to performance degradation (Sections~\ref{sec:Dynamic:ExperimentalResults} and~\ref{sec:CompileTime:ExperimentalResults}).
\end{itemize}

The paper is structured as follows. Section~\ref{sec:Background} covers the IM architecture, with a particular focus on Knights Landing processors, and provides an overview of the current work. Section~\ref{sec:ReverseEngineering} details the reverse engineering process that leads to the discovery of the memory-to-CHA mapping. Sections~\ref{sec:RuntimeOptimization} and~\ref{sec:CompileTime} detail runtime- and compile-time-based approaches, respectively, to optimize coherence traffic and summarize the results of the experimental evaluation phase. Section~\ref{sec:Discussion} discusses the obtained results and related work. Finally, Section~\ref{sec:Concluding} concludes the paper.

\section{Background and Overview}
\label{sec:Background}

\begin{figure}
	\centering
	\includegraphics[width=.8\columnwidth]{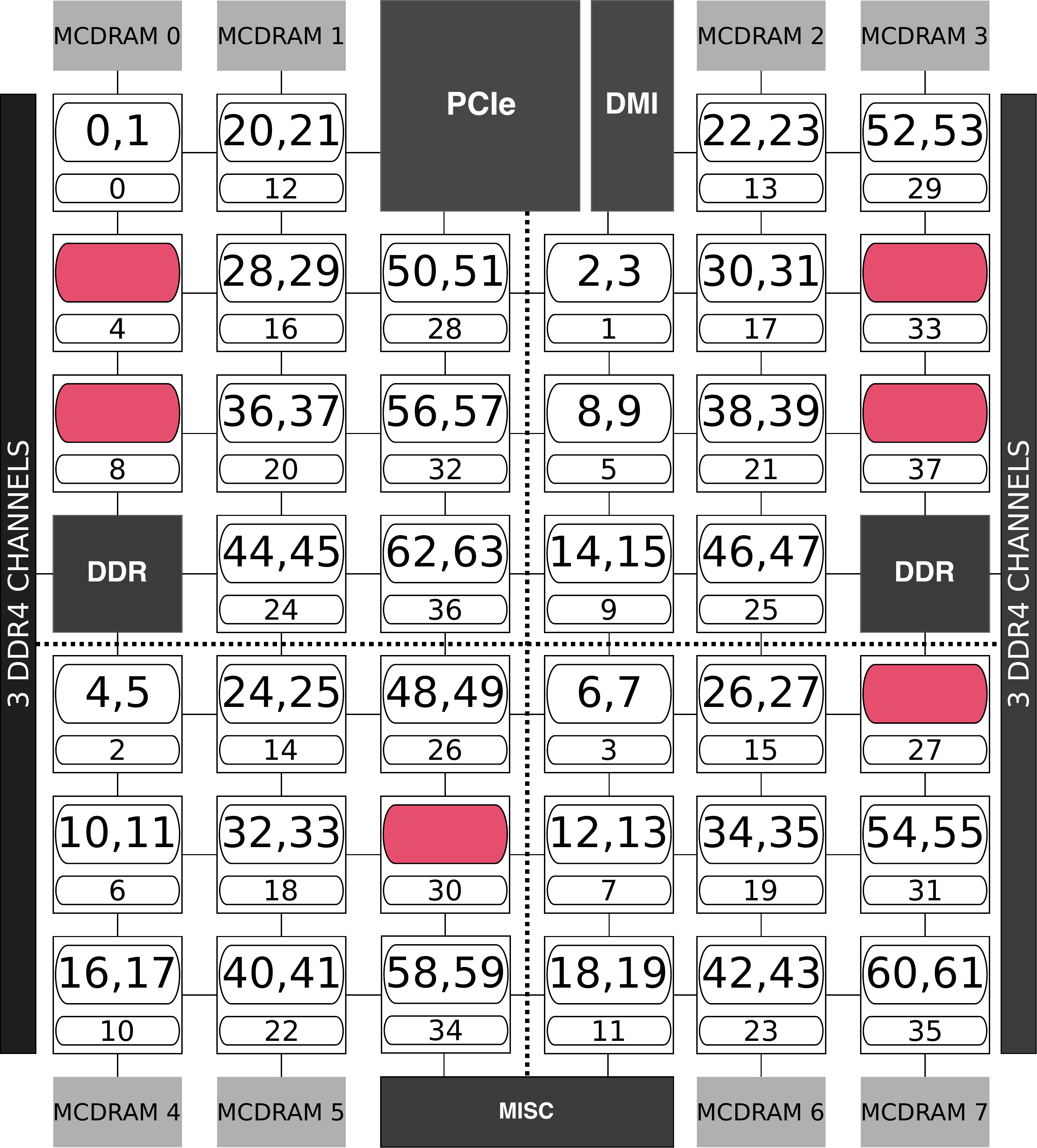}
	\caption{Physical location of logical entities on the KNL floorplan. Each tile contains two cores with the logical IDs shown in the enclosing rectangle. The smaller number below shows the logical ID of each CHA module. Tiles with blank boxes have their cores disabled. A simple naming algorithm can be inferred: CHAs are labeled in sequence, and the set of IDs in each quadrant yield the same value modulo 4. Logical processor IDs are also assigned sequentially, but processors with disabled cores are skipped. More details on the analysis of the logical layout of the processor can be found in~\cite{Marcos:DAC19}.}
	\label{fig:Background:KNL-Floorplan}
\end{figure}

This paper studies the Intel Knights Landing (KNL) architecture as a paramount example of the Intel Mesh interconnect. KNL~\cite{Sodani:KNL} is a manycore processor, including from 64 to 72 cores inside a single die. The processor layout consists of a 2D mesh topology containing 38 tiles, detailed in Fig.~\ref{fig:Background:KNL-Floorplan}. Internally, each tile contains two cores, each with its private L1 instruction and data caches (32~KiB each); and a unified L2 cache (1~MiB) shared among the local cores, but private to the tile.

The KNL processor has two different types of DRAM memory. A Multi-Channel DRAM (MCDRAM) provides high bandwidth through eight interfaces in the corners of the mesh. Besides, two DDR controllers on opposite parts of the chip control three memory channels each. The MCDRAM memory has higher latency than DDR (it is approximately 10\% slower), but the eight interfaces can be accessed simultaneously, providing a much higher bandwidth. 

Messages traverse the mesh using a simple YX routing protocol: a packet always travels vertically first, until it hits its target row. Then, it begins traveling horizontally until it reaches its destination. Each vertical hop takes 1 clock cycle, while horizontal hops take 2 cycles. The mesh features 4 parallel networks, each customized for delivering different types of packets.

KNL employs a directory-based cache coherence mechanism using Intel MESIF~\cite{Goodman:MESIF}, a variant of MESI. In order to alleviate the bottleneck of centralized directories, it features a distributed system in which each tile includes a Caching/Home Agent (CHA) in charge of managing a portion of the directory. Each time a core requests a memory block that does not reside in the local tile caches, the distributed directory is queried. A message is sent to the appropriate CHA (message (1) in Fig.~\ref{fig:Intro:KNL-Latencies}). If the block already resides in one of the L2 caches in the mesh in Forward state\footnote{A cache containing a block in Forward state is in charge of serving said block upon a request. The requestor acquires the block in Forward state, while the sender changes it to Shared.}, the CHA will forward the request to the owner, which will send the data to the requestor in turn (messages (2) and (3) in the figure). In other cases, the data must be fetched from the appropriate memory interface. The data flow shown in the figure exemplifies one of the performance hazards inherent to the KNL architecture: although the data for the requested block lies in the forwarder tile \textbf{F}, just above the requestor \textbf{R}, the coherence data is stored far away in tile \textbf{C}. As it is, 18 cycles are required to transfer the data (10 vertical and 4 horizontal hops). But, if the directory information were stored either in the requestor or in the forwarder, the round trip time of data packets would be of only 2 cycles (2 vertical hops on the mesh). 


The KNL processor features specialized \emph{sub-NUMA} clustering modes, which provide lower memory latency to NUMA-aware applications only (e.g., MPI codes). In this work we focus on the more general \emph{Quadrant} configuration mode, which is the de-facto standard in which any processor can access any memory block. Commonly, the access time of a core to any memory block is assumed to be UMA when in Quadrant mode~\cite{Sodani:KNL, Sabela:IPDPS17}. This is a reasonable assumption, given that memory blocks will be uniformly interleaved across the CHAs and memory interfaces using an opaque, pseudo-random hash function. As a result, the access latency will be averaged out over a sufficient number of accesses for all cores. If the fine-grained behavior of each core is analyzed, the access latency for different memory blocks is, however, not uniform. \citet{Marcos:DAC19} measured the access latencies in Figure~\ref{fig:Intro:KNL-Latencies}, showing that the actual communication costs from different cores to a fixed memory block are far from UMA. More precisely, the coherence traffic causes a systematic degradation of the theoretical optimal memory performance which, on average, creates the illusion of UMA behavior. They further identified the physical placement of logical entities on the processor, shown on Figure~\ref{fig:Background:KNL-Floorplan}, which allows to optimize data and process placement to minimize traffic latencies. Finally, Horro et al. generated the map of the correspondence between each single block of the 16~GiB high-bandwidth MCDRAM memory to its corresponding CHA, by leveraging the performance counters provided by the architecture. Using this map, an inspector-executor approach was proposed to dynamically schedule tasks in irregular codes to processors improving both coherence and data traffic.

Inspector-executor approaches present undesirable runtime overheads. If a closed form of the memory-to-CHA mapping function were known it could be exploited to devise dynamic approaches to task scheduling with lighter overhead, or provided to an optimizing compiler that generated ad-hoc schedules taking advantage of the access latency information. The following section details how the actual pseudo-random memory block mapping over the CHAs was analyzed to extract mapping functions that can be used to predict the CHA assigned to a given physical memory block. This information is then exploited in Sections~\ref{sec:RuntimeOptimization} and~\ref{sec:CompileTime} to devise the proposed optimizations.

\section{Reverse engineering the CHA Mapping}
In hardware designs, pseudo-random mappings often make use of XOR gates, such 
as with Cyclic Redundancy Codes (CRCs), Linear Feedback 
Shift Registers, and other XOR hashes \cite{LFSR:Krawczyk}. XOR mappings can be efficiently implemented in gates 
relative to other forms of pseudo-random mapping binary addresses, such as 
modulo arithmetic of the form $x = (n_1\text{addr} + n_2) \text{ mod } n_3$.

Since full 64-byte cache lines are stored when a CHA location is determined for the data, the address-to-CHA mapping does not make use of address bits 5:0. This section describes the analysis of the mapping data generated by ~\citet{Marcos:DAC19} in order to generate the closed forms for the mapping functions. Table~\ref{tab:addr2CHA} shows the CHA mapping for the first 128 cache lines out of the 256 million mapped locations, i.e., the entire MCDRAM address space.

\newcommand{\ffbox}[1]{%
	{
		\setlength{\fboxsep}{-2\fboxrule}
		\fbox{\hspace{1.2pt}\strut#1\hspace{1.2pt}}
	}
}
\begin{table}[h!tb]
  \sethlcolor{black}
  \caption{Address-to-CHA mapping for the first 128 CHA values out of 256 million. To aid in visualizing, CHA 0 and 1 are shown \ffbox{\textbf{in a box}} and CHA 37 is shown in~{\colorbox{black}{\color{white}white over black}}.}
  \label{tab:addr2CHA}
  \small
  \centering
  \setlength\tabcolsep{2pt}
  \begin{tabular}{@{}r|rrrrrrrrrrrrrrrr@{}}
  \hline
   & \multicolumn{16}{c}{Address bits 9:6} \\
           &  0 &  0 &  0 &  0 &  0 &  0 &  0 &  0 &  1 &  1 &  1 &  1 &  1 &  1 &  1 &  1 \\
   Address &  0 &  0 &  0 &  0 &  1 &  1 &  1 &  1 &  0 &  0 &  0 &  0 &  1 &  1 &  1 &  1 \\
   bits    &  0 &  0 &  1 &  1 &  0 &  0 &  1 &  1 &  0 &  0 &  1 &  1 &  0 &  0 &  1 &  1 \\
   12:10   &  0 &  1 &  0 &  1 &  0 &  1 &  0 &  1 &  0 &  1 &  0 &  1 &  0 &  1 &  0 &  1 \\
  \hline
       000 & 26 &  9 & 24 & 11 & 21 &  6 & 23 &  4 & 31 & 12 & 29 & 14 & 16 &  3 & 18 &  \ffbox{\textbf{1}} \\
       001 & 27 &  8 & 25 & 10 & 20 &  7 & 22 &  5 & 14 & \textcolor{white}{\hl{37}} & 28 & 15 & 33 & 34 & 19 &  \ffbox{\textbf{0}} \\
       010 & 10 & 25 &  8 & 27 &  5 & 22 &  7 & 20 & 15 & 36 & 13 & 30 & 32 & 35 &  2 & 17 \\
       011 & 11 & 24 &  9 & 26 &  4 & 23 &  6 & 21 & 30 & \textcolor{white}{\hl{37}} & 12 & 31 & 33 & 34 &  3 & 16 \\
       100 & 12 & 31 & 14 & 29 &  3 & 16 &  \ffbox{\textbf{1}} & 18 &  9 & 26 & 11 & 24 &  6 & 21 &  4 & 23 \\
       101 & 13 & 30 & 15 & 28 &  2 & 17 &  \ffbox{\textbf{0}} & 19 &  8 & 27 & 34 & 33 &  7 & 20 & \textcolor{white}{\hl{37}} &  6 \\
       110 & 28 & 15 & 30 & 13 & 19 &  \ffbox{\textbf{0}} & 17 &  2 & 25 & 10 & 35 & 32 & 22 &  5 & 36 &  7 \\
       111 & 29 & 14 & 31 & 12 & 18 &  \ffbox{\textbf{1}} & 16 &  3 & 24 & 11 & 34 & 33 & 23 &  4 & \textcolor{white}{\hl{37}} & 22 \\
  \hline
  \end{tabular}
\end{table}

Given values for CHA from 0 to 37, 6 bits are needed to represent this number, 
but given that 38 is not a power of 2, we did not expect to see a 
straightforward XOR equation of address bits for each CHA bit. However, given 
the ease of computing binary functions in hardware, we did expect and found 
that each bit for the CHA value can be computed independently (again, as 
opposed to a scheme like $addr \text{ mod } 38$). The process we follow to 
determine the equations for CHA bits 0 and 1 is:

\begin{itemize}
\item Search for CHA bit functions of the form $f=a_1 \oplus a_2 \oplus ... \oplus a_{n-1} \oplus z(a_n,a_{n+1},...)$
\item Analyze CHA bit toggle rates for each address bit, determining direct XOR behaviors where toggling a single bit in the address also toggles the CHA bit as shown in Table~\ref{tab:a6to34}.
\item Analyze the limited input binary function of the bits not directly used 
as XOR values as shown in Table~\ref{tab:a33to29}.
\end{itemize}

For instance, consider the toggle frequency for $\text{CHA}_0$ when different bits of the address are toggled shown in Table~\ref{tab:a6to34}. As can be seen, 99.93\% of the time toggling $a_6$ or $a_8$ changes the result of $\text{CHA}_0$, whereas toggling $a_7$ almost never affects its value. It can be concluded that the mapping function for bit 0 must be of the form $\text{CHA}_0 = a_6 \oplus a_8 \oplus ...$, where ``$...$'' is yet to be determined. The fact that the data is not 100\% conclusive is attributable to measurement errors for a minority of cases in the performance counter-based mapping process.

The analysis of the toggle frequency finds that some of the bits $A_{28:6}$ are directly XOR'd into $\text{CHA}_0$, while some others do not appear at all. However, the study also shows that bits $A_{33:29}$ affect the function, but not in the same categorical way. The toggle frequency is somewhere between 1\% and 95\%. In order to reverse engineer the role of these bits in the function, the limited input binary function of $A{33:29}$ is analyzed to detect which combinations of these bits toggle the result of the partial XOR function built from $A_{28:6}$. This reverse engineering process yields the functions $\text{CHA}_0$ and $\text{CHA}_1$ in Figure~\ref{fig:RevEng:CHAEqs} for the 2 least significant bits of the CHA.

\begin{table}[h!tb]
  \caption{$\text{CHA}_0$ toggle frequency when toggling address bits $a_6$ to $a_{34}$. In this case, values greater than 0.99 or less than 0.01 indicate errors in the CHA predicted based on performance counters, and are interpreted as 1 and 0, respectively.}
  \label{tab:a6to34}
  \small
  \centering
  \setlength\tabcolsep{2pt}
  \begin{tabular}{r@{\hspace{3ex}}|@{\hspace{3ex}}c@{\hspace{3ex}}c}
  \hline
   Address & Fraction of time $\text{CHA}_0$ & Address role \\
   Bit     & toggles when $a_n$ does & in $\text{CHA}_0$ equation \\
   \hline
   $a_6$   & 0.9993 & XOR \\
   $a_7$   & 0.0001 & Ignore \\
   $a_8$   & 0.9993 & XOR \\
   $a_9$   & 0.9994 & XOR \\
   $a_{10}$   & 0.9994 & XOR \\
   $a_{11}$   & 0.0002 & Ignore \\
    ... & ... & ... \\
   $a_{29}$   & 0.0120 & Function \\
   $a_{30}$   & 0.9458 & Function \\
   $a_{31}$   & 0.9444 & Function \\
   $a_{32}$   & 0.0555 & Function \\
   $a_{33}$   & 0.0546 & Function \\
   $a_{34}$   & 0.0000 & Ignore \\
  \hline
  \end{tabular}
\end{table}

\begin{table}[h!tb]
  \caption{For $\text{CHA}_0$, bits 29 to 33 do not directly get XOR'd with other bits, but are part of a function that itself is XOR'd with those bits.}
  \label{tab:a33to29}
  \small
  \centering
  \setlength\tabcolsep{2pt}
  \begin{tabular}{r@{\hspace{3ex}}|@{\hspace{3ex}}c}
  \hline
   Address      & Fraction of time $\text{CHA}_0$ is low when\\
   Bits 33:29   & result of direct XOR bits is high \\
   \hline
   00000 & 0.0 \\
   00001 & 0.0038 \\
   00010 & 0.0 \\
   00011 & 0.0 \\
   00100 & 0.0 \\
   00101 & 0.0 \\
   00110 & 1.0 \\
    ... & ... \\
   11010 & 0.0 \\
   11011 & 0.0 \\
   11100 & 0.0 \\
   11101 & 0.001 \\
   11110 & 0.9983 \\
   11111 & 1.0 \\
  \hline
  \end{tabular}
\end{table}

\begin{figure*}
	\begin{flalign*}
		\text{CHA}_{0} = & a_{6} \oplus a_{8} \oplus a_{9} \oplus a_{10} \oplus a_{14} \oplus a_{15} \oplus a_{17} \oplus a_{18} \oplus a_{20} \oplus a_{23} \oplus a_{27} \oplus ((a_{30}  a_{31}) | ( \overline{a_{30}}  \overline{a_{31}}(a_{32} | a_{33})))\\
		\text{CHA}_{1} = & a_{6} \oplus  a_{7} \oplus  a_{8} \oplus  a_{12} \oplus  a_{16} \oplus  a_{17} \oplus  a_{20} \oplus  a_{21} \oplus  a_{22} \oplus  a_{23} \oplus  a_{24} \oplus  a_{25} \oplus  a_{26} \oplus  a_{28} \oplus a_{30} \oplus  a_{33}\\
		\text{CHA}_{2} = & f \overline{g}\text{, where:}&\\	
		& f =  a_{8} \oplus  a_{9} \oplus  a_{12} \oplus  a_{15} \oplus  a_{16} \oplus  a_{18} \oplus  a_{21} \oplus  a_{22} \oplus  a_{23} \oplus  a_{25} \oplus  a_{26} \oplus  a_{28} \oplus \overline{a_{30}}(a_{31} | a_{32} | a_{33}))\\
		&g = ((a_{11} \oplus  a_{16} \oplus  a_{17} \oplus  a_{21} \oplus  a_{23} \oplus  a_{26} \oplus  a_{27} \oplus  a_{28} \oplus  a_{29} \oplus  a_{31}) | (a_{10} \oplus  a_{15} \oplus  a_{16} \oplus  a_{20} \oplus  a_{22} \oplus a_{25} \oplus a_{26} \oplus\\
		&a_{27} \oplus  a_{28} \oplus  \overline{a_{30}} \oplus  a_{34})) (a_{6} \oplus  a_{12} \oplus  a_{20} \oplus  a_{21} \oplus  a_{22} \oplus  a_{23} \oplus  a_{28} \oplus  a_{32} \oplus  a_{34}) (a_{7} \oplus a_{12} \oplus a_{14} \oplus  a_{17} \oplus  a_{18} \oplus  a_{19} \oplus\\
		&a_{22} \oplus  a_{23} \oplus  a_{25} \oplus  a_{26} \oplus  a_{27} \oplus  a_{28} \oplus  a_{29} \oplus  a_{32} \oplus  a_{34}) (a_{9} \oplus  a_{14} \oplus  a_{15} \oplus a_{19} \oplus  a_{21} \oplus  a_{24} \oplus  a_{25} \oplus  a_{26} \oplus  a_{27} \oplus\\
		&a_{29} \oplus  a_{33} \oplus  a_{34}) (a_{13} \oplus  a_{14} \oplus  a_{18} \oplus  a_{24} \oplus  a_{26} \oplus  a_{28} \oplus  a_{29} \oplus  a_{31} \oplus  a_{33} \oplus  a_{34})\\
		\text{CHA}_{3} = & f  \overline{g} | h\text{, where:}&\\
		& f = a_{8} \oplus  a_{13} \oplus  a_{14} \oplus  a_{18} \oplus  a_{20} \oplus  a_{23} \oplus  a_{24} \oplus  a_{25} \oplus  a_{26} \oplus  a_{28} \oplus (a_{30} | a_{31} | a_{32}) (a_{32} \oplus \overline{a_{33}})\\
		& g = ((a_{11} \oplus  a_{16} \oplus  a_{17} \oplus  a_{21} \oplus  a_{23} \oplus  a_{26} \oplus  a_{27} \oplus  a_{28} \oplus  a_{29} \oplus  a_{31}) | (a_{10} \oplus  a_{15} \oplus  a_{16} \oplus  a_{20} \oplus  a_{22} \oplus  a_{25} \oplus  a_{26} \oplus\\
		&a_{27} \oplus  a_{28} \oplus  \overline{a_{30}} \oplus  a_{34})) (\overline{a_{7}} \oplus  a_{12} \oplus  a_{13} \oplus  a_{17} \oplus  a_{19} \oplus  a_{22} \oplus  a_{23} \oplus  a_{24} \oplus  a_{25} \oplus  a_{27} \oplus  a_{31} \oplus  a_{32} \oplus  a_{33}) (a_{9} \oplus  a_{14} \oplus\\
		&a_{15} \oplus  a_{19} \oplus  a_{21} \oplus  a_{24} \oplus  a_{25} \oplus  a_{26} \oplus  a_{27} \oplus  a_{29} \oplus  a_{33} \oplus  a_{34})\\
		& h = ((a_{11} \oplus  a_{16} \oplus  a_{17} \oplus  a_{21} \oplus  a_{23} \oplus  a_{26} \oplus  a_{27} \oplus  a_{28} \oplus  a_{29} \oplus  a_{31}) | (a_{10} \oplus  a_{15} \oplus  a_{16} \oplus  a_{20} \oplus  a_{22} \oplus  a_{25} \oplus  a_{26} \oplus\\
		&a_{27} \oplus a_{28} \oplus  \overline{a_{30}} \oplus  a_{34})) (\overline{a_{6}} \oplus  a_{13} \oplus  a_{15} \oplus  a_{16} \oplus  a_{17} \oplus  a_{18} \oplus  a_{20} \oplus  a_{21} \oplus  a_{26} \oplus  a_{29} \oplus  a_{34}) (\overline{a_{7}} \oplus  a_{13} \oplus  a_{14} \oplus  a_{15} \oplus\\
		&a_{16} \oplus  a_{19} \oplus  a_{25} \oplus  a_{27} \oplus  a_{34}) (\overline{a_{8}} \oplus  a_{15} \oplus  a_{17} \oplus  a_{18} \oplus  a_{19} \oplus  a_{27} \oplus  a_{28} \oplus  a_{29} \oplus  \overline{a_{30}} \oplus  a_{31} \oplus  a_{32} \oplus  a_{34}) (a_{9} \oplus  a_{14} \oplus\\
		&a_{15} \oplus  a_{19} \oplus  a_{21} \oplus  a_{24} \oplus  a_{25} \oplus  a_{26} \oplus  a_{27} \oplus  a_{29} \oplus  a_{33} \oplus  a_{34}) (\overline{a_{12}} \oplus  a_{14} \oplus  a_{15} \oplus  a_{16} \oplus  a_{17} \oplus  a_{22} \oplus  a_{23} \oplus  a_{24} \oplus\\
		&a_{31} \oplus  a_{32} \oplus  a_{33} \oplus  a_{34})\\
		\text{CHA}_{4} = & f  \overline{g} | g  h\text{, where:}&\\
		& f = a_{6} \oplus  a_{11} \oplus  a_{12} \oplus  a_{16} \oplus  a_{18} \oplus  a_{21} \oplus  a_{22} \oplus  a_{23} \oplus  a_{24} \oplus  a_{26} \oplus  a_{30} \oplus  a_{31} \oplus  a_{32}\\
		& g = ((a_{11} \oplus  a_{16} \oplus  a_{17} \oplus  a_{21} \oplus  a_{23} \oplus  a_{26} \oplus  a_{27} \oplus  a_{28} \oplus  a_{29} \oplus  a_{31}) | (a_{10} \oplus  a_{15} \oplus  a_{16} \oplus  a_{20} \oplus  a_{22} \oplus  a_{25} \oplus  a_{26} \oplus\\
		&a_{27} \oplus  a_{28} \oplus  \overline{a_{30}} \oplus  a_{34})) (\overline{a_{7}} \oplus  a_{12} \oplus  a_{13} \oplus  a_{17} \oplus  a_{19} \oplus  a_{22} \oplus  a_{23} \oplus  a_{24} \oplus  a_{25} \oplus  a_{27} \oplus  a_{31} \oplus  a_{32} \oplus  a_{33}) (a_{9} \oplus  a_{14} \oplus\\
		&a_{15} \oplus  a_{19} \oplus  a_{21} \oplus  a_{24} \oplus  a_{25} \oplus  a_{26} \oplus  a_{27} \oplus  a_{29} \oplus  a_{33} \oplus  a_{34})\\
		&h = (\overline{a_{10}} \oplus  a_{11} \oplus  a_{13} \oplus  a_{16} \oplus  a_{17} \oplus  a_{18} \oplus  a_{19} \oplus  a_{20} \oplus  a_{21} \oplus  a_{22} \oplus  a_{27} \oplus  a_{28} \oplus  \overline{a_{30}} \oplus  a_{31} \oplus  a_{33} \oplus  a_{34}) (\overline{a_{6}} \oplus  a_{12} \oplus\\
		&a_{13} \oplus  a_{14} \oplus  a_{18} \oplus  a_{20} \oplus  a_{21} \oplus  a_{22} \oplus  a_{23} \oplus  a_{24} \oplus  a_{26} \oplus  a_{29} \oplus  a_{31} \oplus  a_{32} \oplus  a_{33}) (\overline{a_{7}} \oplus  a_{12} \oplus  a_{13} \oplus  a_{17} \oplus  a_{19} \oplus  a_{22} \oplus\\
		&a_{23} \oplus  a_{24} \oplus  a_{25} \oplus  a_{27} \oplus  a_{31} \oplus  a_{32} \oplus  a_{33}) (\overline{a_{8}} \oplus  a_{12} \oplus  a_{14} \oplus  a_{16} \oplus  a_{18} \oplus  a_{19} \oplus  a_{22} \oplus  a_{23} \oplus  a_{24} \oplus  a_{27} \oplus  a_{28} \oplus  a_{29} \oplus\\
		&\overline{a_{30}} \oplus  a_{33}) (a_{9} \oplus  a_{14} \oplus  a_{15} \oplus  a_{19} \oplus  a_{21} \oplus  a_{24} \oplus  a_{25} \oplus  a_{26} \oplus  a_{27} \oplus  a_{29} \oplus  a_{33} \oplus  a_{34})\\
		\text{CHA}_{5} = & f  \overline{g}\text{, where:}&\\
		& f = ((a_{10} \oplus  a_{15} \oplus  a_{16} \oplus  a_{20} \oplus  a_{22} \oplus  a_{25} \oplus  a_{26} \oplus  a_{27} \oplus  a_{28} \oplus  \overline{a_{30}} \oplus  a_{34}) | (a_{11} \oplus  a_{16} \oplus  a_{17} \oplus  a_{21} \oplus  a_{23} \oplus  a_{26} \oplus\\
		&a_{27} \oplus  a_{28} \oplus  a_{29} \oplus  a_{31})) (\overline{a_{7}} \oplus  a_{12} \oplus  a_{13} \oplus  a_{17} \oplus  a_{19} \oplus  a_{22} \oplus  a_{23} \oplus  a_{24} \oplus  a_{25} \oplus  a_{27} \oplus  a_{31} \oplus  a_{32} \oplus  a_{33}) (a_{9} \oplus  a_{14} \oplus\\
		&a_{15} \oplus  a_{19} \oplus  a_{21} \oplus  a_{24} \oplus  a_{25} \oplus  a_{26} \oplus  a_{27} \oplus  a_{29} \oplus  a_{33} \oplus  a_{34})\\
		&g = (\overline{a_{6}} \oplus  a_{12} \oplus  a_{13} \oplus  a_{14} \oplus  a_{18} \oplus  a_{20} \oplus  a_{21} \oplus  a_{22} \oplus  a_{23} \oplus  a_{24} \oplus  a_{26} \oplus  a_{29} \oplus  a_{31} \oplus  a_{32} \oplus  a_{33}) (\overline{a_{8}} \oplus  a_{12} \oplus  a_{14} \oplus\\
		&a_{16} \oplus  a_{18} \oplus  a_{19} \oplus  a_{22} \oplus  a_{23} \oplus  a_{24} \oplus  a_{27} \oplus  a_{28} \oplus  a_{29} \oplus  \overline{a_{30}} \oplus  a_{33})\\
	\end{flalign*}
	
	\label{fig:RevEng:CHAEqs}
	\caption{Reverse-engineered mapping function between memory blocks and CHAs.}
\end{figure*}

Although the number of CHA locations (38) is not divisible by 4, we found that 
CHA bits 0 and 1 are each on for 50\% of the addresses, and as seen in 
Figure~\ref{fig:Background:KNL-Floorplan} this distributes data evenly among 
the 4 quadrants of the die. $\text{CHA}_1=1$ indicates the data is in the lower 
half of the die; $\text{CHA}_0=1$ indicates the data is on the right side of 
the die. However, CHA bits 2 through 5 have more complex functions in order to 
reasonably distribute data among the 38 CHA values. The employed 
approach was as follows:

\begin{itemize}
\item For CHA bits 2 through 5, search for CHA bit functions of the form $f\overline{g}|h$ or $f\overline{g}|gh$, where $f$, $g$, and $h$ may be primarily XOR-based but may include other values. We found that all bits have a base pattern function $f$, bits 2 through 5 have a masking function $g$, and bits 3 and 4 have a masking function $h$.
\item Analyze the CHA bit toggle rate for each address bit, indicating XOR behaviors within the base $f$ function.
\item Analyze the limited input binary function of the bits not directly used as XOR values within $f$ to complete the initial estimate of $f$ function.
\item Using the baseline $f$ function to predict CHA bit values, compare the expected CHA value in the performance counter-based mapping with the resulting $f$ to build $g$ and $h$ bit by bit, starting at the least significant bit. 
\end{itemize}

As an example, we will describe the process of determining the $g$ function for $\text{CHA}_2$ with reference to Table~\ref{tab:CHA2}. The $f$ function includes $a_8 \oplus a_9 \oplus a_{12}$ which results in regular blocks of 1's and 0's when the address bits 13:6 are varied with a total of 128 1's and 128 0's in the set of 256 cache lines. However, the performance counter data implies that 6 of those 1's are actually 0's. Note that in the binary representation of 0 through 37, bit 2 is high 18/38 = 47\% of the time, hence the simple 50/50 XOR equation from $f$ needs to be masked to 0 in some pseudo-random locations resulting on $\text{CHA}_2 = f\overline{g}$. Given where the masking occurs in Table~\ref{tab:CHA2}, we surmise the structure of the $g$ function to be $((a_{11} \oplus ...)|(a_{10} \oplus ...))(a_6 \oplus ...)(a_7 \oplus ...)(a_9 \oplus ...)(...)$ where ``$...$'' represents unknown functions of higher order bits. By comparing $f\overline{g}$ to the known CHA locations with partially completed $g$ functions, we build up the complete mask functions detailed in Figure~\ref{fig:RevEng:CHAEqs}.

%
%

\begin{table}[h!tb]
  \caption{$\text{CHA}_2$ for the first 256 cache lines. Given a base function which includes $a_8 \oplus a_9 \oplus a_{12}$, the 6 boxed \ffbox{\textbf{0}} positions show where a masking function is used}.
  \label{tab:CHA2}
  \small
  \centering
  \setlength\tabcolsep{2pt}
  \begin{tabular}{@{}r|rrrrrrrrrrrrrrrr@{}}
  \hline
   & \multicolumn{16}{c}{Address bits 9:6} \\
           &  0 &  0 &  0 &  0 &  0 &  0 &  0 &  0 &  1 &  1 &  1 &  1 &  1 &  1 &  1 &  1 \\
   Address &  0 &  0 &  0 &  0 &  1 &  1 &  1 &  1 &  0 &  0 &  0 &  0 &  1 &  1 &  1 &  1 \\
   bits    &  0 &  0 &  1 &  1 &  0 &  0 &  1 &  1 &  0 &  0 &  1 &  1 &  0 &  0 &  1 &  1 \\
   13:10   &  0 &  1 &  0 &  1 &  0 &  1 &  0 &  1 &  0 &  1 &  0 &  1 &  0 &  1 &  0 &  1 \\
  \hline
      0000 &  0 &  0 &  0 &  0 &  1 &  1 &  1 &  1 &  1 &  1 &  1 &  1 &  0 &  0 &  0 &  0 \\
      0001 &  0 &  0 &  0 &  0 &  1 &  1 &  1 &  1 &  1 &  1 &  1 &  1 &  0 &  0 &  0 &  0 \\
      0010 &  0 &  0 &  0 &  0 &  1 &  1 &  1 &  1 &  1 &  1 &  1 &  1 &  0 &  0 &  0 &  0 \\
      0011 &  0 &  0 &  0 &  0 &  1 &  1 &  1 &  1 &  1 &  1 &  1 &  1 &  0 &  0 &  0 &  0 \\
      0100 &  1 &  1 &  1 &  1 &  0 &  0 &  0 &  0 &  0 &  0 &  0 &  0 &  1 &  1 &  1 &  1 \\
      0101 &  1 &  1 &  1 &  1 &  0 &  0 &  0 &  0 &  0 &  0 &  0 &  0 &  1 &  1 &  1 &  1 \\
      0110 &  1 &  1 &  1 &  1 &  0 &  0 &  0 &  0 &  0 &  0 &  0 &  0 &  1 &  1 &  1 &  1 \\
      0111 &  1 &  1 &  1 &  1 &  0 &  0 &  0 &  0 &  0 &  0 &  0 &  0 &  1 &  1 &  1 &  1 \\
      1000 &  0 &  0 &  0 &  0 &  1 &  1 &  1 &  1 &  1 &  1 &  1 &  1 &  0 &  0 &  0 &  0 \\
      1001 &  0 &  0 &  0 &  0 &  1 &  1 &  1 &  1 &  1 &  1 &  1 &  \ffbox{\textbf{0}} &  0 &  0 &  0 &  0 \\
      1010 &  0 &  0 &  0 &  0 &  1 &  1 &  1 &  1 &  1 &  1 &  1 &  \ffbox{\textbf{0}} &  0 &  0 &  0 &  0 \\
      1011 &  0 &  0 &  0 &  0 &  1 &  1 &  1 &  1 &  1 &  1 &  1 &  \ffbox{\textbf{0}} &  0 &  0 &  0 &  0 \\
      1100 &  1 &  1 &  1 &  1 &  0 &  0 &  0 &  0 &  0 &  0 &  0 &  0 &  1 &  1 &  1 &  1 \\
      1101 &  1 &  1 &  1 &  1 &  0 &  0 &  0 &  0 &  0 &  0 &  0 &  0 &  \ffbox{\textbf{0}} &  1 &  1 &  1 \\
      1110 &  1 &  1 &  1 &  1 &  0 &  0 &  0 &  0 &  0 &  0 &  0 &  0 &  \ffbox{\textbf{0}} &  1 &  1 &  1 \\
      1111 &  1 &  1 &  1 &  1 &  0 &  0 &  0 &  0 &  0 &  0 &  0 &  0 &  \ffbox{\textbf{0}} &  1 &  1 &  1 \\

  \hline
  \end{tabular}
\end{table}

Like $\text{CHA}_2$, $\text{CHA}_5$ uses a base $f$ function and a mask-to-0 $g$ function. Bits $\text{CHA}_3$ and $\text{CHA}_4$ had a base $f$ function, mask-to-0 $g$ function, and mask-to-1 $h$ function. The $h$ function was found by recognizing where the $f\overline{g}$ pattern itself was not producing correct predictions. The process of determining the $f$ function by observing XOR toggle indications and solving the 4- or 5-bit binary function remaining can be automated. In theory, given a rough constraint on the types of functions to be considered, the process of finding the $g$ and $h$ functions could also be automated by searching for mispredictions of the $f$ function to the true result. However, future architectures may vary the mapping structure so our process of involving a human to interpret binary results and build the equations may remain common for this type of task.

\label{sec:ReverseEngineering}

\section{Runtime optimization}
\label{sec:RuntimeOptimization}


%

As mentioned in Section~\ref{sec:Background}, \citet{Marcos:DAC19} developed an inspector-executor approach to the optimization of coherence traffic in KNL processors. This approach is limited to irregular codes, and consists in transforming the data layout so that the data to be accessed by each tile lie in memory blocks for which the coherence information was assigned to nearby CHAs. This approach has an important overhead during the inspection phase. First, the input data need to be physically copied to target memory blocks with the required coherence properties. Then, the associated indirection arrays need to be recomputed. Lastly, the resulting data are now spread across a much larger region of memory, in order to find suitable memory blocks, and therefore cache locality is degraded and the number of page faults increased. With the closed form of the mapping functions exposed in Section~\ref{sec:ReverseEngineering}, it is possible to apply this approach to general codes, instead of being restricted to irregular computations. The basic idea is to encode the schedule of tasks not on the indirection arrays, but to exploit the properties of the mapping function.



Consider the general matrix-vector multiplication code depicted in Figure~\ref{fig:Dynamic:Gesummv}. This is an interesting problem because of its simplicity, its transversality, and because of the fact that it is memory-bound in modern processors. As such, it will benefit from increasing the memory throughput. The dominant part of the memory footprint of the computation is the access to matrix $B$, and therefore the following analysis will be centered on trying to optimize its access.

\newsavebox{\gesummv}
\begin{lrbox}{\gesummv}
	\begin{lstlisting}
#pragma omp parallel for
for (int i = 0; i < N; ++i)
	for (int j = 0; j < N; ++j)
		y[i] += B[i * N + j] * x[j];
	\end{lstlisting}
\end{lrbox}

\begin{figure}[h]
	\centering
	\subfloat{\usebox{\gesummv}}
	\caption{Scalar code for general matrix-vector multiplication parallelized using a static block schedule.}
	\label{fig:Dynamic:Gesummv}
\end{figure}

Given the complexity of the mapping functions, it is implausible to dynamically perform a very fine-grained scheduling of iterations to tiles that will actually have the required coherence information in its local CHA. Besides, this would imbalance the computation, as some CHAs have up to 20\% more memory blocks than others due to the irregular nature of the mapping functions. Instead, we focus on the quadrant granularity, emulating the behavior of the sub-NUMA modes of the machine by ensuring that each tile computes data with coherence information resident on its quadrant only. The approach followed for scheduling iterations in this fashion is described in the following.


The quadrant mapping benefits from a convenient feature of the address-to-CHA functions. As noted in Section~\ref{sec:ReverseEngineering}, and due to the physical placement of logical CHAs on the network-on-chip shown in Figure~\ref{fig:Background:KNL-Floorplan}, bits $\text{CHA}_0 = c_0$ and $\text{CHA}_1 = c_1$ identify the quadrant $c_1c_0$ in which the CHA is located. Consider the $k-th$ memory block with address $A^k$ aligned to a 256-byte boundary, i.e., $k$ is a multiple of 4. Bits $A^k_{5:0}$ express an offset inside the memory block, and therefore are not used in the computation of the associated CHA. Because of the 256-byte alignment, $A^k_{7:6} = 00_b$. The address of the next memory block, $A^{k+1} = A^k + 64$, will share its most significant bits with $A^k$, i.e., $A^{k+1}_{63:8} = A^k_{63:8}$, and $A^{k+1}_{7:6} = 01_b$. Since $A_6$ participates in the XOR computation in the equations for $\text{CHA}_1$ and $\text{CHA}_0$ in Figure~\ref{fig:RevEng:CHAEqs}, it can be determined that the least significant bits of its associated CHA will be flipped, i.e., if the associated quadrant for $A^k$ is $c_1c_0$ then the associated quadrant for $A^{k+1}$ will be $\overline{c_1}~\overline{c_0}$. Similarly, $A^{k+2}_{7:6} = 10_b$ and its associated quadrant will be $\overline{c_1}c_0$; and $A^{k+3}_{7:6} = 11_b$ and its associated quadrant is $c_0\overline{c_1}$. This results in the convenient organization that precisely 1 out of every 4 cache lines is in each physical quadrant, allowing parallel access routines to evenly divide up work among physical processors. 

In the proposed sub-NUMA schedule a processor located in quadrant $c_1c_0$ will process only memory blocks with associated CHA in the same quadrant. After processing a block at address $A$, the next address in the same quadrant could be located at $A+100_b$, $A+101_b$, $A+110_b$, or $A+111_b$ depending on $A_{33:8}$. Determining which of the 4 addresses is next in our quadrant mathematically requires to compute the full CHA equations discovered in Section~\ref{sec:ReverseEngineering}. However, these are complex so these computations should be performed as little as possible. The actual offset required to compute the next address in quadrant $c_1c_0$ has a fixed pattern for address bits $A_{12:8}$, which allows a 64-bit register to store the offsets for the next 32 cache lines. In this way, processors stepping through memory can thus avoid full computation of the mapping function 31 out of each 32 iterations.

\subsection{Experimental results}
\label{sec:Dynamic:ExperimentalResults}

In order to have full control over the executed instructions, the original code from Figure~\ref{fig:Dynamic:Gesummv} is manually vectorized using AVX-512 intrinsics as shown in Figure~\ref{fig:Dynamic:GesummvVect}. In this way, opaque optimizations that may bias the comparison of different schedules are avoided. This section focuses on single-precision floating point arithmetic only, but all obtained results are directly extrapolable to double precision FP.

\newsavebox{\gesummvvect}
\begin{lrbox}{\gesummvvect}
\begin{lstlisting}
#pragma omp parallel for
for (int i = 0; i < N; ++i) {
	_m512 bb, bx, accum;
	accum0 = _mm512_setzero_ps();
	for (int j = 0; j < N; j += 16) {
		bb    = _mm512_load_ps(&B[i * N + j]);
		bx    = _mm512_load_ps(&x[j]);
		accum = _mm512_fmadd_ps(bb, bx, accum);
	}
	y[i] = _mm512_reduce_add_ps(accum);
}
\end{lstlisting}
\end{lrbox}

\begin{figure}[h]
	\centering
	\subfloat{\usebox{\gesummvvect}}
	\caption{Manually vectorized code for general matrix-vector multiplication 
	parallelized using a static block schedule.}
	\label{fig:Dynamic:GesummvVect}
\end{figure}

Both the code in Figure~\ref{fig:Dynamic:GesummvVect} and the equivalent sub-NUMA schedule are executed on an Intel x200 7210 running at the base frequency of 1.30 GHz, to avoid turbo-related variations. The codes were compiled using ICC 19.1.1.217, with flags \texttt{-Ofast -xKNL -qopenmp}. They are executed on 64 threads using \texttt{KMP\_AFFINITY=scatter}. Heap variables are stored into 1~GiB hugepages via \texttt{hugectl --heap}, and these hugepages are guaranteed to be allocated in the MCDRAM address space using \texttt{numatcl -m 1}. The experiments are run with $N=16384$, which makes matrix $B$ take up 1~GiB of memory, that is, an entire hugepage.

\begin{figure}[h]
	\centering
	\includegraphics[width=.49\textwidth]{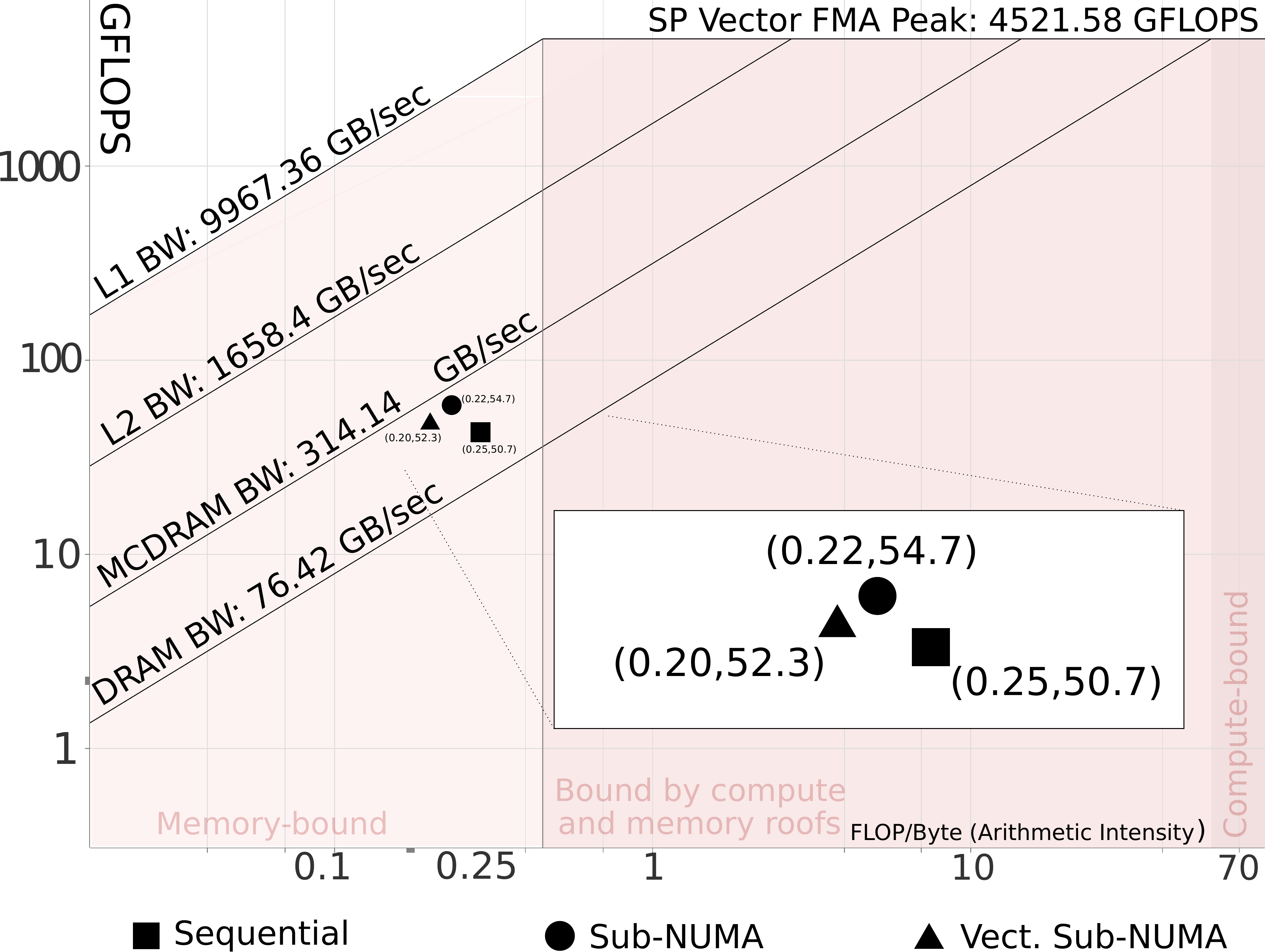}
	\caption{Roofline plot for the matrix-vector multiplication using single precision arithmetic.}
	\label{fig:Dynamic:Roofline}
\end{figure}

The roofline model generated by Intel Advisor~\cite{Intel:Advisor} for these codes is shown in Figure~\ref{fig:Dynamic:Roofline}. For these experiments, the hardware prefetcher was manually turned off using Model Specific Registers (MSR)~\cite{Intel:HWPF} in order to observe the raw effect of the proposed coherence traffic optimizations without interference. As shown in the figure, the sequential schedule achieves 50.7 GFLOPS for an arithmetic intensity (AI) of 0.25, which is approximately 65\% of the roofline for that AI, whereas the sub-NUMA schedule achieves 54.7 GFLOPS for an AI of 0.22, or 81\% of the roofline. The GFLOPS have increased and the AI has decreased, due to the additional memory traffic required to compute the sub-NUMA schedule, resulting in a large net increase of the percentage of peak performance that is obtained. Executions with double-precision arithmetic achieve the same approximate results, but dividing the number of raw GFLOPS by 2.

\begin{figure}[ht!]
	\centering
	\includegraphics[width=\columnwidth]{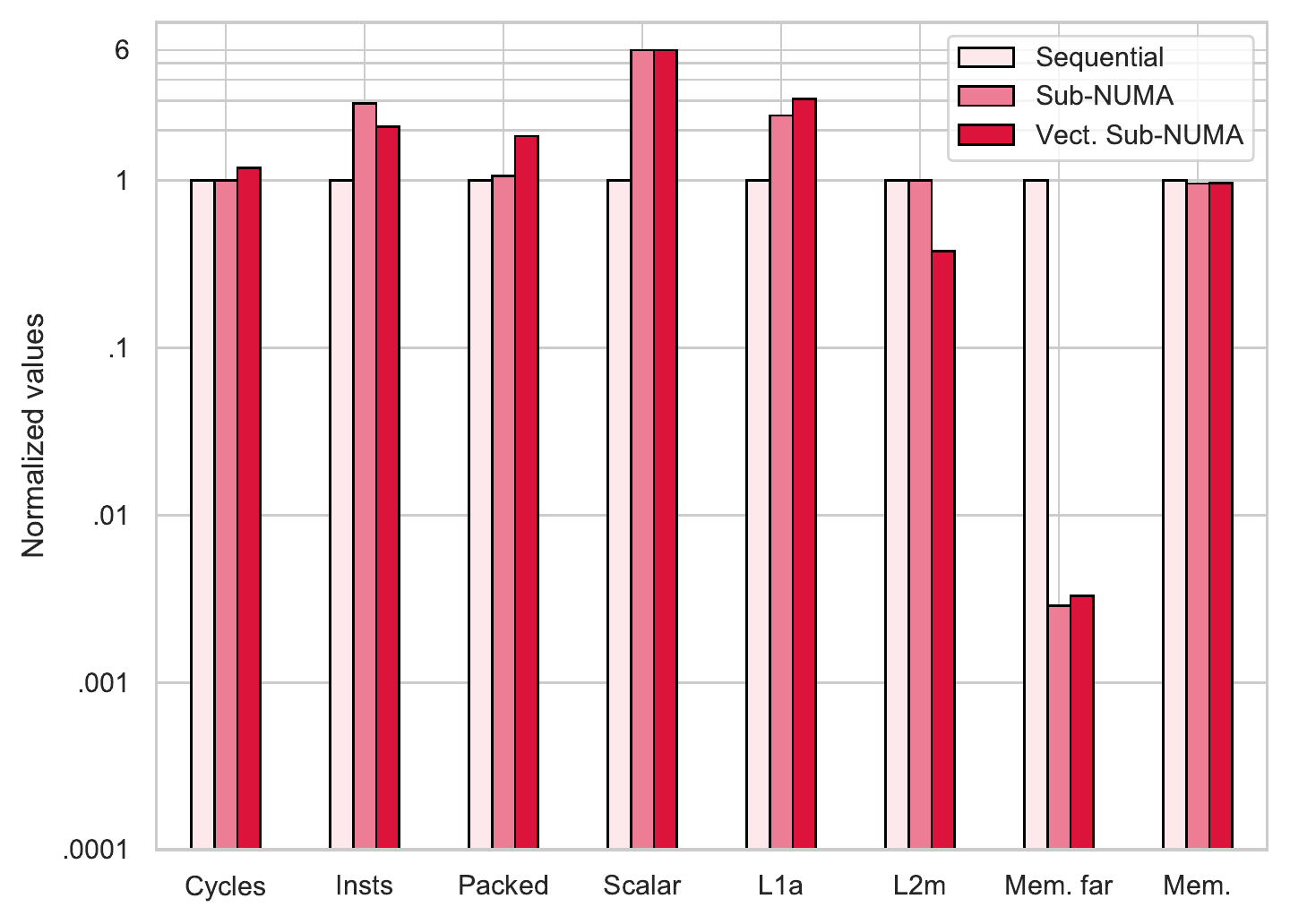}
	\caption{Sum of selected performance counters for all threads. Logarithmic scale is used for the Y axis. The figure shows the number of cycles, instructions issued, packed SIMD instructions, scalar SIMD instructions, L1 data accesses, L2 misses, MCDRAM ``far'' accesses to other quadrants in the NoC, and total number of MCDRAM accesses. Values are normalized to those of the sequential schedule.}
	\label{fig:Dynamic:PAPICounters}
\end{figure}

The improvement in raw performance measured by the roofline model, however, can be deceitful. Although the sub-NUMA schedule achieves a higher FLOP count, it also executes additional instructions on non-consecutive memory blocks, causing a degradation in cache behavior and ultimately execution time. In order to more closely investigate the effect of the proposed optimization, selected performance counters were measured for several different execution setups. The results are shown in Figure~\ref{fig:Dynamic:PAPICounters}. In order to compute the sub-NUMA schedule, the number of instructions to be executed almost triples, increasing by 188\%. The largest share of these are data L1 loads and stores, which grow by 145\%. This increase, however, is absorbed by the L2 cache, and the L2 misses remain virtually identical. There is a very significant increase in the IPC of these codes, which goes from 18.6 in the original version to 53.15 in the sub-NUMA schedule. The memory latency, approximated by the \texttt{OFFCORE\_RESPONSE\_0:OUTSTANDING} performance counter, is slightly decreased by 1.8\%. All these variables compound for an almost zero net effect on execution time: execution cycles are reduced by a modest 0.8\%.

In order to try to decrease the schedule-related computations, a modified version which employs vectorization operations for offset computation was developed. In essence, the offsets for each 32 consecutive memory blocks are now computed using AVX-512 arithmetic. This version, labeled ``Vect. sub-NUMA'' in Figures~\ref{fig:Dynamic:Roofline} and~\ref{fig:Dynamic:PAPICounters} achieves to reduce the number of instructions by 37.8\% with respect to the regular sub-NUMA schedule. However, it worsens register pressure, increasing L1 accesses by a further 26\%. As a result, the GFLOPS decrease to 52.3, and so does the AI to 0.20, for a grand total of 82.4\% of the peak performance.

%



As previously mentioned, these results were executed after disabling the hardware prefetching. The reason is that the sub-NUMA schedule does not access memory sequentially, and is at a tremendous disadvantage against the sequential schedule when the prefetcher is enabled, which would absorb and eliminate any potential advantage from the sub-NUMA schedule. In fact, when enabling the hardware prefetcher the performance of the sequential schedule is improved by 1.2x, whereas it is detrimental for sub-NUMA (i.e., its performance slightly decreases by approximately 5\%) as it features a pseudo-random access pattern that mimics the memory-to-CHA mapping funtions.

\section{Compile-time optimization}
\label{sec:CompileTime}

As shown by the experiments in the previous section,  improving the mesh locality during runtime has an important impact on other execution parameters due to the pseudo-random nature of the memory-to-CHA mapping functions and their computational complexity. A different way to exploit this knowledge is to optimize the scheduling of completely static codes during the compilation stage.

\newsavebox{\exfig}
\begin{lrbox}{\exfig}
	\includegraphics[width=0.5\textwidth]{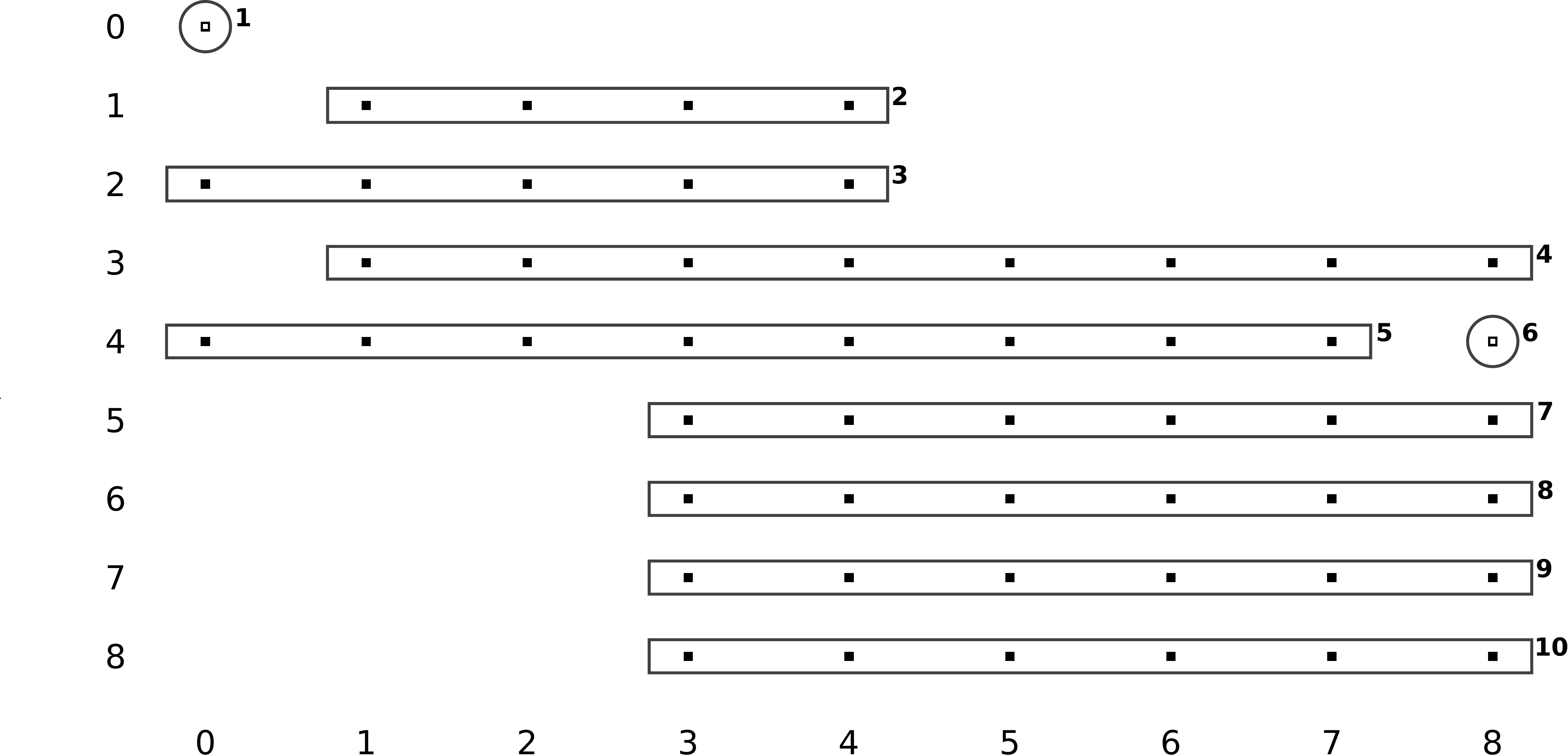}
\end{lrbox}

\newsavebox{\excode}
\begin{lrbox}{\excode}
	\begin{lstlisting}
	y[0] += A[0] * x[0];
	y[1] += A[1:4] * x[1:4];
	y[2] += A[5:9] * x[0:4];
	y[3] += A[10:17] * x[1:8];
	y[4] += A[18:25] * x[0:7];
	y[4] += A[26] * x[8];
	y[5] += A[27:32] * x[3:8];
	y[6] += A[33:38] * x[3:8];
	y[7] += A[39:44] * x[3:8];
	y[8] += A[45:50] * x[3:8];
	\end{lstlisting}
\end{lrbox}

\begin{figure*}
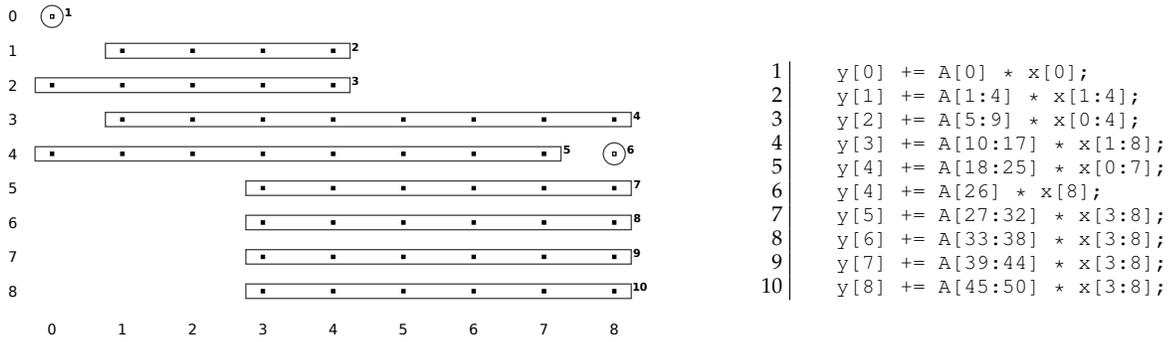

	\subfloat{\usebox{\exfig}}
	\qquad\qquad\qquad
	\subfloat{\vbox to .85\ht\exfig{\usebox\excode}}
	\vspace{-1.25cm}
	\caption{Sets of regular subcomputations built for the Sparse Matrix-Vector multiplication of matrix \texttt{FIDAP/ex7} in the SuiteSparse repository. The figure in the left shows the location of the nonzero points in the upper left corner of the matrix. Each identified regular subcomputation is marked as a rectangle enclosing several nonzeros, and captured as an AVX-512 operation, as shown in the pseudo-code on the right.}
	\label{fig:CompileTime:ex7}
\end{figure*}

\citet{Gabriel:PLDI19} recently proposed a data-specific code generation technique for the optimization of sparse-immutable codes, including artificial neural network inference. In essence, this approach automatically builds sets of regular subcomputations by mining for regular subregions in the irregular data structure. The resulting code is specialized to the sparsity structure of the input matrix, but does not employ indirection arrays, improving predictability and SIMD vectorizability. This section focuses on the sparse matrix-vector multiplication (SpMV) as an immediate target of this class of data-specific optimizations.

A graphical depiction of a small subset of operations performed by the sparse matrix-vector multiplication of matrix \texttt{FIDAP/ex7}, included in the SuiteSparse Matrix Collection~\cite{UF-Collection} is offered in Figure~\ref{fig:CompileTime:ex7}. For many sparse matrices, this code generation approach delivers better performance than the generic, irregular alternative. Besides promoting vectorization, data-specific approaches encode the matrix structure implicitly in the program source. This does not only reduce the number of memory accesses, but collaterally stores the matrix structure in the first-level instruction cache, which is classically underutilized for small irregular codes such as SpMV. The effect is similar to extending the first-level data cache: matrix structure will be stored in the instruction cache (since it is embedded in the code), whereas actual matrix values will be stored in the data cache. The immediate disadvantage is that the code grows proportionally to the matrix size. Still, for sufficiently regular sparse matrices the combined size for structure and data values (the program footprint) will be small enough as to benefit from this tradeoff.

As opposed to the dynamic approach of Section~\ref{sec:RuntimeOptimization}, the static optimization has no explicit execution overhead. As such, the schedule of each computation can be carefully analyzed and planned in order to improve coherence traffic. Note that, as opposed to the dynamic approach in which the mapping functions could be applied on already-allocated memory, in this case the memory allocation must be statically known. The approach employed for this is detailed in Section~\ref{sec:CompileTime:PhysicalAddresses}. For the remainder of this section it is assumed that the physical address associated to each data block in the program is statically known.

Consider the generic SpMV statement $s$ executed by the data-specific approach:
\[ s : y_i = A_j \cdot x_k \]
\noindent{}Note that this statement does not include irregular indices, since the code has been generated for a specific input matrix with a fixed sparsity structure, as exemplified in Figure~\ref{fig:CompileTime:ex7}. Consequently, the compiler has static knowledge of all the memory movements that will be required for executing each specific part of the code. At a glance, the proposed compile-time approach computes an access cost for each statement in the data-specific SpMV code for each tile in the processor, and then schedules operations across the mesh following a greedy approach. Access costs are dynamically updated during the scheduling process to reflect the updated placement of each memory block in the private caches of each tile.

\begin{figure}
	\centering
	\includegraphics[width=.8\columnwidth]{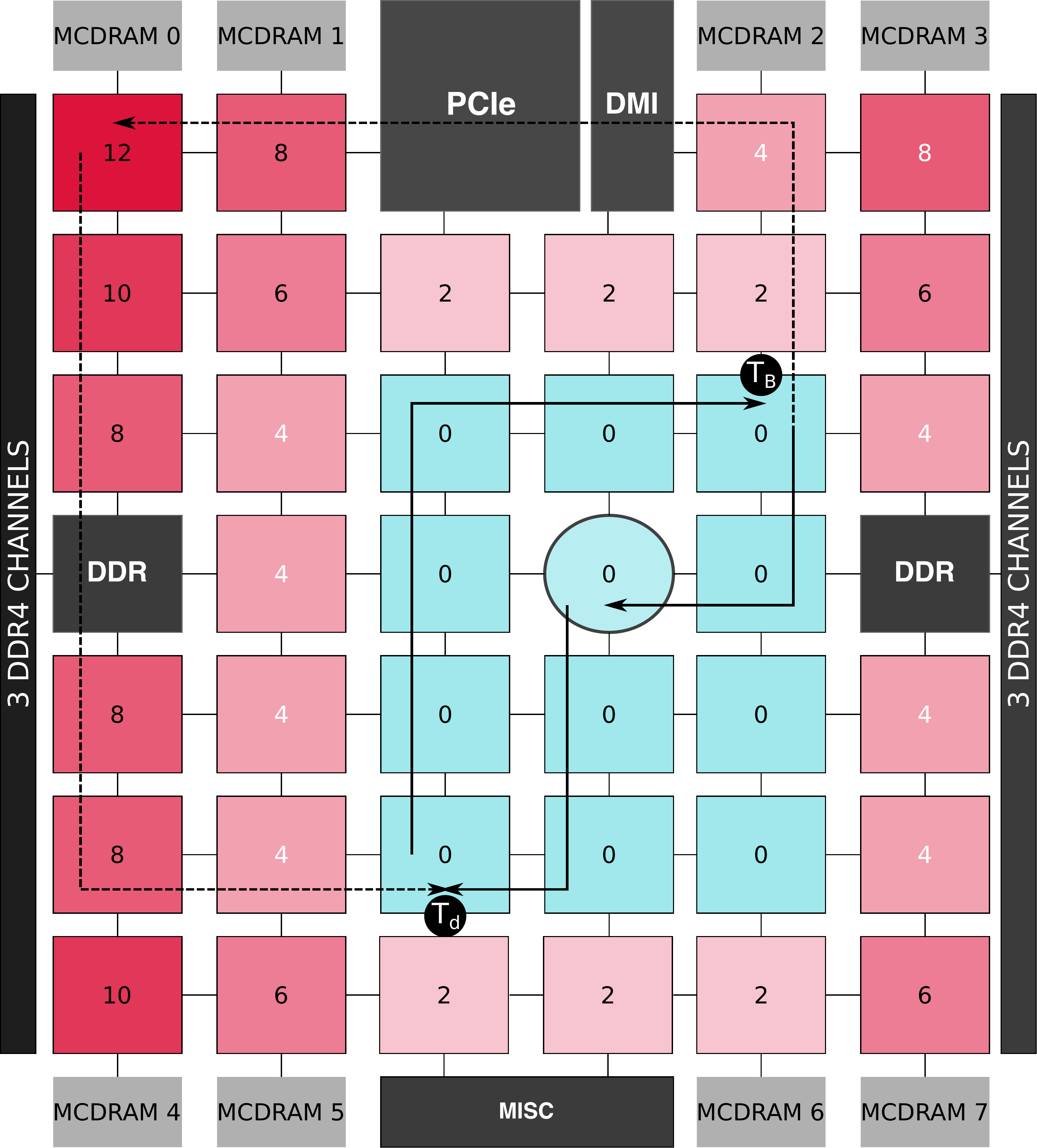}
	\caption{Overhead, in mesh cycles, of accessing a block of data resident in the L2 cache of tile $T_B$, and with coherence information resident in tile $T_d$. All the tiles inside the rectangle defined by $T_d$ and $T_B$ access data in 14 cycles with zero associated overhead. In any communication, first the CHA at $T_d$ is queried. Then the CHA sends a \emph{F}orward request to the source L2 cache at $T_B$, which sends the data back to the requestor. For any tile inside the 0-overhead rectangle, being closer to the CHA means a shorter travel time for the query, and a larger one for the response. These compensate one another, yielding zero net effect. For tiles outside the rectangle, the overhead is compounded by the extra time that the query needs to enter the rectangle, plus the extra time that the response needs to arrive back at the requestor from inside a rectangle tile.}
	\label{fig:CompileTime:MeshRectangle}
\end{figure}

Consider a data block $B$ with directory information associated to tile $T_d$ and actual data accessed through tile $T_B$. The actual source of data can either be the private L2 cache of tile $T_B$, if the associated tile is the $F$orwarder for $B$; or $T_B$ can be one of the tiles with an associated memory interface, which will serve $B$ after reading it from memory. Regardless of the actual coherence status of $B$, in order to access the data the requestor tile will send a message to $T_d$, which will forward the request to $T_B$, which in turn will send $B$ back to the requestor. Figure~\ref{fig:CompileTime:MeshRectangle} illustrates this situation. Note that $T_d$ and $T_B$ constitute the opposite corners of a rectangle on the network-on-chip (NoC) which contains the tiles that can access $B$ with minimum latency. Tiles outside this rectangle incur extra latency, which can be computed as $2 \times (2 \times D_x + D_y)$, where $D_x$ and $D_y$ are the horizontal and vertical distances from the tile to the rectangle, respectively.

Based on these access times, a scheduling system is developed, conceptually described in Algorithm~\ref{alg:CompileTime:Scheduling}. Each tile in the NoC is visited in order, and for each of them the subset of operations to be executed on that tile is selected in a greedy, iterative fashion, choosing the one with the smallest data movement cost at each iteration, until that tile reaches its balanced load. The cost $\tau$ of executing each statement $s$ in tile $t$ is computed as:
\[ \tau(s,t) = \tau( y_i, t ) + \tau( A_j, t ) + \tau( x_k, t ) \]
\noindent{}That is, the aggregated cost of accessing memory blocks $y_i$, $A_j$, and $x_k$ from tile $t$. For each individual memory block, its access cost is computed as:
\[ \tau( B, t ) = \lambda_B + 2 \times \left( 2 \times D_x(t,R_B) + D_y(t,R_B) \right) \]
\noindent{}where:
\begin{itemize}
	\item $\lambda_B$ is a factor that depends on the latency to physically access $B$, including $12$ cycles 
	for accessing a private L2 in the NoC, and $115$ cycles for accessing an MCDRAM interface.
	\item $D_{x/y}(t,R_B)$ is the horizontal/vertical distance from the requestor tile $t$ to the rectangle defined by the tiles in its opposite corners $T_B$ and $T_d$, containing the data and the coherence information, respectively, as described in Figure~\ref{fig:CompileTime:MeshRectangle}.
\end{itemize}

The order in which each of the tiles is visited is carefully selected: those with worst-case trip times are selected first. For instance, the upper-left tile in the NoC has a worst-case round trip time of 32 cycles when accessing data with $T_d$ or $T_B$ on the bottom-right tile. However, the round trip time from a central tile to any other tile in the mesh is of at most 18 cycles.

\begin{algorithm}
	\SetAlgoLined
	\KwIn{Set $\mathcal{S}$ of SpMV statements to be scheduled}
	\KwIn{Set $\mathcal{T}$ of tiles in the NoC}
	\KwOut{Schedule $\Theta( \mathcal{S} ) \rightarrow \mathcal{T}$}
	Compute $L_T$ = total number of FLOPS in $\mathcal{S}$\;
	Compute $L_b$ = $\frac{L_T}{\#\mathcal{T}}$\, the number of FLOPs to be computed by each tile to balance load\;
	\ForEach{tile $t \in \mathcal{T}$}{%
		\While{$Load(t) < L_b$}{%
			Select $s \in \mathcal{S} : \tau( s, t ) \leq \tau( s', t ), \forall s, s' \in \mathcal{S}$\;
			Assign $\Theta(s) = t$\;
			Update $\mathcal{S} = \mathcal{S} - \{s\}$\;
		}
	}
	\caption{Static scheduling of SpMV operations}
	\label{alg:CompileTime:Scheduling}
\end{algorithm}

Note that the schedules generated by this static optimization process are no longer sub-NUMA, as opposed to the dynamic approach in Section~\ref{sec:RuntimeOptimization}. In this case, there is no runtime constraint enforcing quick computation of the schedule, so the system can use the full fine-grained information about memory-to-CHA mapping to decide whether accessing data on a different quadrant will be the best option from a coherence traffic point of view.

Once all operations are scheduled, the code is generated specifically for each tile. In order to reduce code sizes, affine compression may be applied to group similar operations together on regular affine loops~\cite{Gabriel:TC19}. These do not employ indirection arrays, being still fully vectorizable, while reducing the pressure on the instruction cache.

\subsection{Fixing Physical Addresses}
\label{sec:CompileTime:PhysicalAddresses}

One of the challenges of static scheduling with this class of pseudo-random functions is that it is not possible to compute the associated CHA of a virtual address, as the 34 least-significant bits of the address will be used. Even with 1~GiB hugepage sizes, the maximum supported by the architecture, only 30 bits remain unchanged during the virtual-to-physical address translation. This means that the code cannot rely simply on page alignment, as can be done for cache optimization, and must target specific physical pages.

In order to fix the physical pages that are assigned to a specific application, we employ 1~GiB hugepages. Since the MCDRAM address space has only 16~GiB in total, there will only be 16 possible pages that can be assigned to our application. The assignment order varies slightly depending on the machine state upon launch. To overcome this difficulty we employ a hybrid static/dynamic approach. During the static analysis, the code generated assumes that specific 1~GiB hugepages will be allocated to the different data structures in the program. These assumptions are registered in static constant variables in the source code. During runtime, an executor overallocates as many 1~GiB pages as possible. Then, it translates their virtual addresses to physical addresses by reading the process pagemap in \texttt{/proc}. Finally, it assigns the required hugepages to the data structures in the code by comparing the allocated physical addresses to the static constant variables assumed during the scheduling process, and frees the remaining, unused ones.

\subsection{Experimental Results}
\label{sec:CompileTime:ExperimentalResults}

We generate data-specific codes for more than 20 sparse matrices selected from the range of matrices between 1 million and 10 million nonzeros in the SuiteSparse repository. The upper bound is used for tractability purposes. The lower bound to ensure sufficiently large operation. The selected matrices were the cluster centroids resulting from running k-means on SuiteSparse and using regularity and size as the target characteristics \cite{Gabriel:PLDI19}. Each of the selected matrices was processed to extract the data-specific operations required by its SpMV. Three different implementations were generated for testing:

\begin{itemize}
	\item The generic irregular version of Figure~\ref{fig:CompileTime:irregular_spmv}.
	\item A data-specific version with sequential schedule, as described by \citet{Gabriel:PLDI19}.
	\item A data-specific version containing exactly the same set of operations, but scheduled in a coherence-aware fashion using Algorithm~\ref{alg:CompileTime:Scheduling}.
\end{itemize}

\begin{figure}
	\begin{minipage}{\textwidth}
		\begin{lstlisting}
#pragma omp parallel for private(j)
for(i = 0; i < n; ++i) {
	y[i] = 0;
	for( j = pos[i]; j < pos[i+1]; ++j)
	y[i] += A[j] * x[cols[j]];
}
		\end{lstlisting}%
	\end{minipage}%
	\caption{Classical, irregular SpMV code}
	\label{fig:CompileTime:irregular_spmv}
\end{figure}%

Codes are compiled using ICC 19.1.1.217 with \texttt{-Ofast -xKNL -qopenmp}. They are executed on an Intel x200 7210, running at the base frequency of 1.30~GHz, to avoid turbo-related variations, using 64 threads, one per core in the NoC. Ten repetitions were performed for each execution, and average values are reported for each thread after discarding outliers (identified as values $x$ such that $ |x-\bar{X}| > 3\sigma(X)$). For the generic irregular version and the sequentially-scheduled data-specific one the ``scatter'' thread placement is employed. For the coherence-aware version an ad-hoc assignment is employed, ensuring that each thread is executed on the appropriate statically scheduled tile. These codes are typically very large in size, explicitly containing the full set of operations to be performed for multiplying a sparse matrix by a given vector. Executable sizes vary between 39 and 206~MB. As for dynamic scheduling, \texttt{hugectl --heap} and \texttt{numactl -m 1} are used to control the use of hugepages and memory domains. The hardware prefetcher is enabled for all the experiments in this section.

The data-specific versions were found to be 2.1x faster on average than the generic irregular version. This is a clear indication that a manycore architecture with light, vectorization-oriented processors is not well geared towards irregular codes, which feature many control flow-related instructions such as induction variable increments and branches. The data-specific versions perform, on aggregate, 4.7x less L1 accesses, but incur 1.2x more L1 data misses. The L1 instruction misses increase by 39.9x. This increase is mostly absorbed by the L2 cache and the hardware prefetcher, however, and overall the number of L2 misses is only 12.8\% higher in the data-specific versions. Furthermore, these additional misses are resolved locally by the mesh, and the number of MCDRAM accesses decreases by 21.6\%. In summary, the memory behavior, which is potentially the weakest runtime aspect of a data-specific version, is not significantly worsened. In exchange, the data-specific codes execute 5.4x less instructions, including 2.3x less scalar operations and 859x more vector operations. The biggest culprit in runtime difference is precisely the number of executed instructions, and the number of stalls due to missing reservation stations is 4.9x larger on irregular codes. Due to these intrinsic differences in the nature of each implementation, we drop the irregular version of SpMV in the following experiments, and focus on comparing only the sequential and coherence-aware schedules of the data-specific implementations.

\begin{figure}
	\includegraphics[width=\columnwidth]{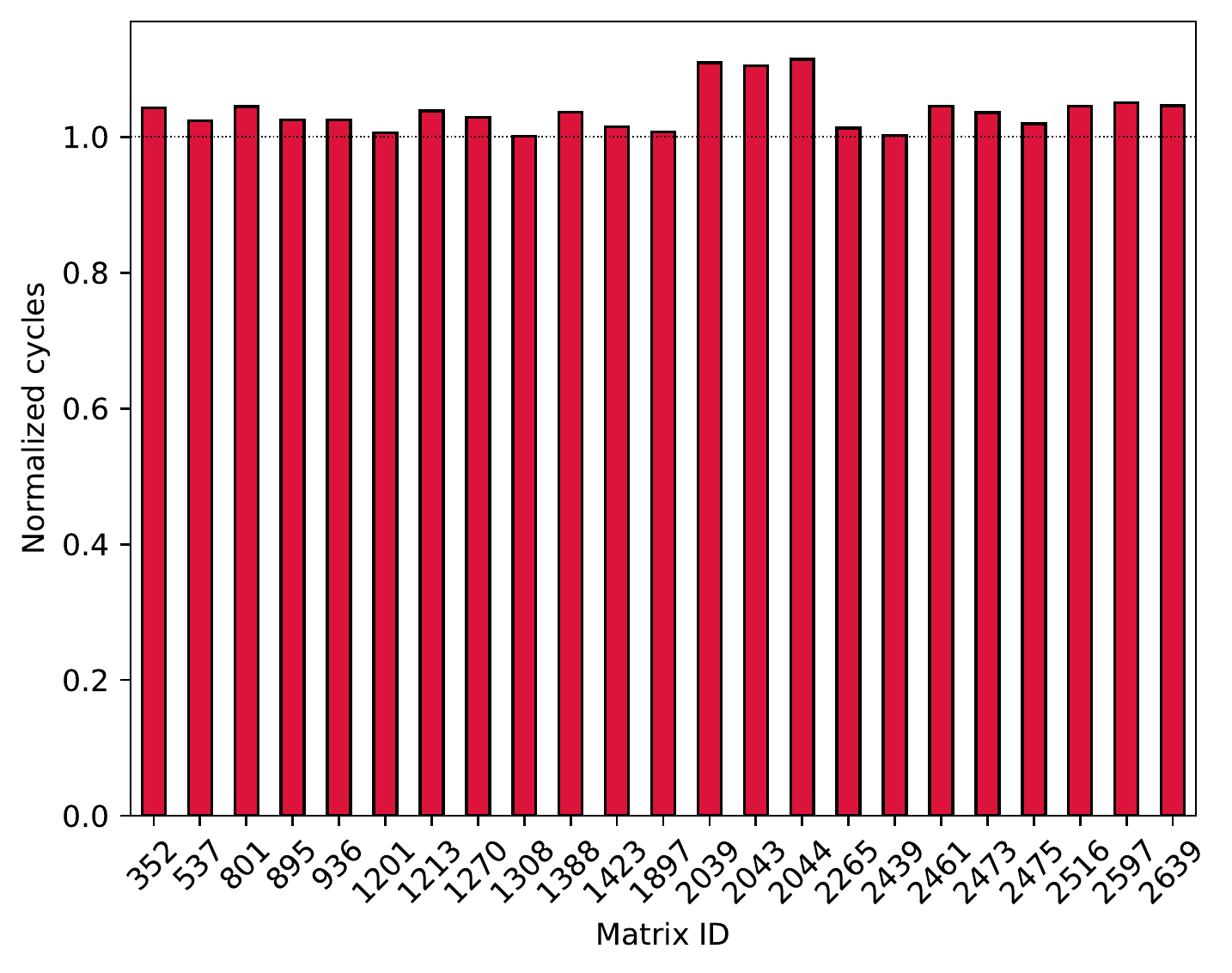}
	\caption{Execution cycles of the coherence-aware schedule, normalized to those of the sequential schedule.}
	\label{fig:CompileTime:UNHALTED_CORE_CYCLES}
\end{figure}

From a performance point of view, on aggregate the coherence-aware schedule increases execution time by 3.2\%. The detailed execution cycles obtained for the SpMV of each matrix are shown in Figure~\ref{fig:CompileTime:UNHALTED_CORE_CYCLES}. None of the matrices achieves a performance improvement, the best one being 0.1\% slower than the baseline. For some matrices the operation is noticeably slower, the extreme case being 10.3\% less performant.

\begin{figure}
	\includegraphics[width=\columnwidth]{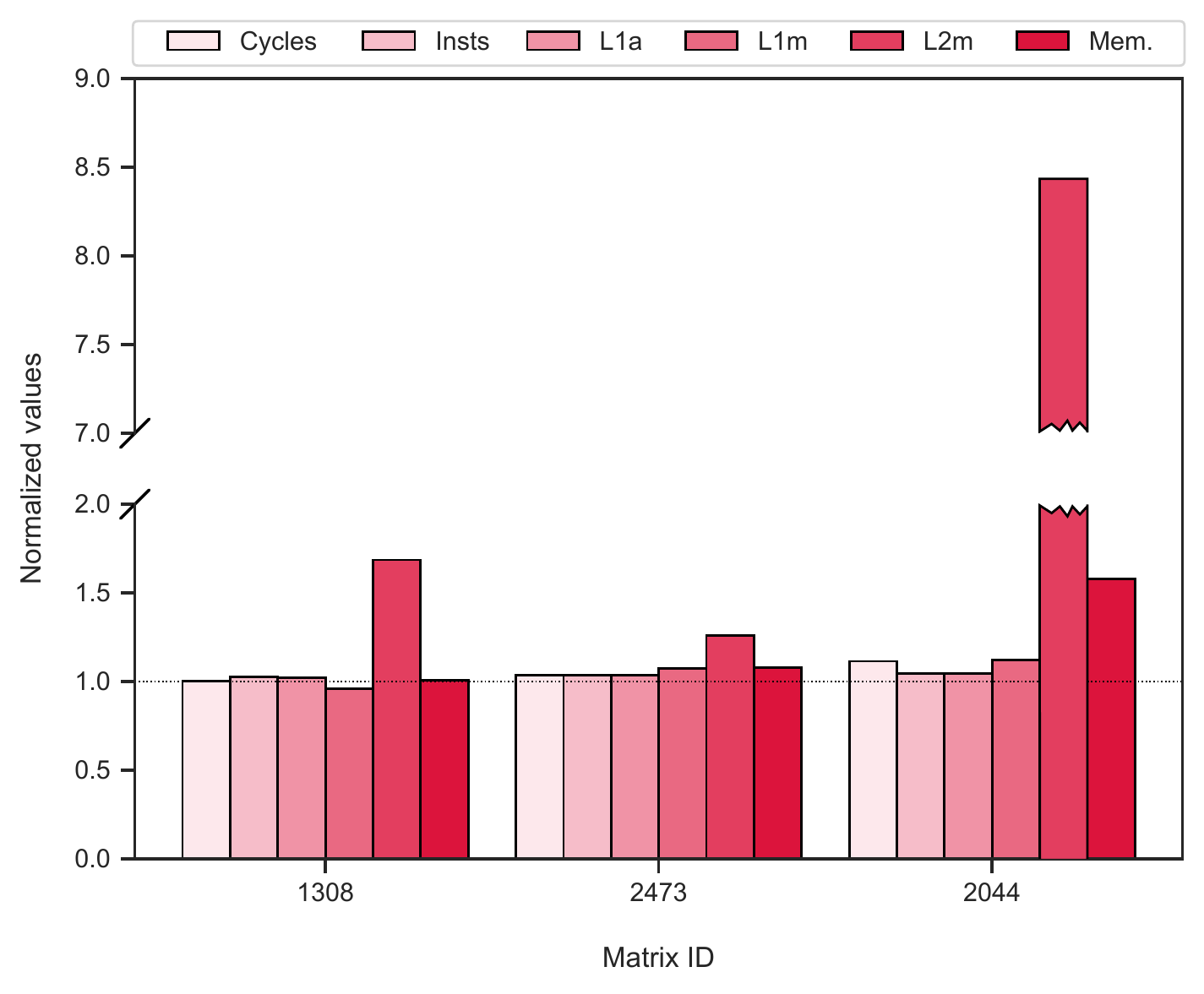}
	\caption{Execution cycles, executed instructions, data L1 accesses, data L1 misses, L2 misses, and MCDRAM accesses of the coherence-aware schedule for selected matrices in the experimental set: the one with the best relative performance (\#1308), the one with the worst one (\#2044), and the middle clase (\#2473). Values are normalized to those of the sequential schedule.}
	\label{fig:CompileTime:PerfCounters}
\end{figure}

\begin{figure}
	\includegraphics[width=\columnwidth]{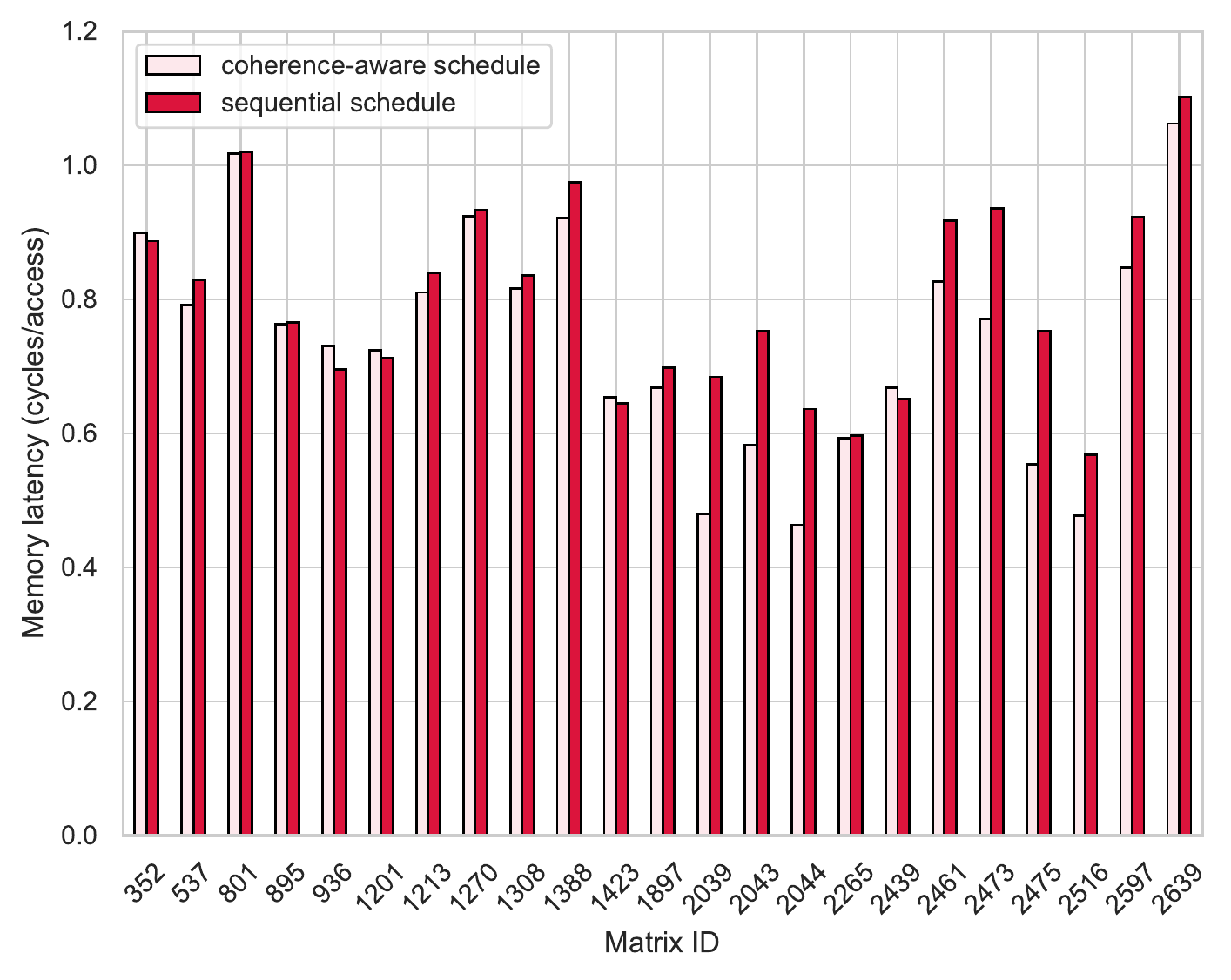}
	\caption{Memory latency of the coherence-aware schedule normalized to the sequential schedule baseline for all the matrices in the experimental set.}
	\label{fig:CompileTime:MemLatCPI}
\end{figure}

In order to study in more detail the reasons for this performance degradation, three selected matrices are closely examined. Selected performance counters for these matrices are detailed in Figure~\ref{fig:CompileTime:PerfCounters}. On careful inspection the performance is strongly correlated with the number of MCDRAM accesses incurred by each version of the code (R=0.91). The conclusion to be inferred from these experiments is that, even with fully static scheduling, CHA locality cannot be appropriately leveraged to improve performance of data-specific sparse codes. The reason is that, due to the pseudo-random nature of the assignment between memory blocks and CHAs, rescheduling the code to promote the access of nearby CHAs to improve the cache coherence traffic patterns necessarily impacts cache locality negatively for codes benefiting from sequential data access. Even though SpMV has a varying degree of randomness in the access to the $x$ vector, the matrix data in $A$ can be accessed sequentially, and this is a huge advantage of the sequential schedule, particularly taking into account the hardware prefetcher. Despite the performance degradation, a careful analysis of the performance counters evidences that the coherence-aware schedule broadly improves memory latency, as shown in Figure~\ref{fig:CompileTime:MemLatCPI}, by 10\% on aggregate. Average latency goes from 0.77 cyles per access in the sequential schedule, to 0.70 cycles per access in the coherence-aware schedule. The IPC is very slightly increased, going from 12.37 to 12.42.

\section{Discussion and Related Work}
\label{sec:Discussion}
When \citet{Marcos:DAC19} initially explored the optimization of coherence traffic on the Knights Landing NoC, they observed a clear effect on the application performance due to affinity relationships between cores and CHAs. This work was based on a pre-computed assignment of memory blocks to CHAs, which takes up 256 MiB for the MCDRAM memory. The optimized scheduling was performed dynamically in an inspector-executor fashion, which represents a very costly step that would negate any actual performance benefit in a real setting. Furthermore, the rescheduling could only be applied to irregular codes.

Based on these promising results, the current work focused on reverse engineering the functions behind this mapping. To our knowledge, this is the first work that has managed to discover this information. The authors expected these functions to be useful to alleviate the overhead of the inspector-executor approach, in addition to being usable by architecture-specific compilers that could perform low-level optimizations of coherence traffic. However, these expectations were toned down by the actual shape of the mapping functions. Although the XOR-based functions are cheap to implement in hardware and widely used for other non-regular mappings, such as the assignment between memory blocks and LLC slices in Intel Core processors~\cite{Farshin:RevEng,Irazoqui:RevEng,Maurice:RevEng}, they are costly to compute in software. Still, this can be overcome if the mapping presents some kind of regularity that can be exploited by carefully optimizing the code and schedules. The coupling of the software complexity of the functions, together with their pseudo-random nature is what ultimately makes it virtually impossible to benefit from coherence traffic improvement in these designs.

For all these inconveniences from the high-performance computing point of view, the approach followed by Intel has many advantages in everyday computing. It is implausible to write a code that systematically accesses only a particular set of CHAs, making them into a bottleneck. Such a bottleneck can happen with regular mappings, such as a modulo-based mapping that can suffer from systematic conflicts for certain access patterns. Furthermore, it manages to distribute memory blocks across the quadrants in the NoC in a fair fashion, ensuring that all of them have to manage the same amount of information on aggregate. This is no simple task, given the irregularity of the NoC, which features a non-power of two number of tiles, unevenly distributed across quadrants. Still, the price to pay is an all-to-all coherence traffic pattern which requires dedicated communication rings to handle.

Going forward, it would be desirable to improve this design, coupling the directory distribution that avoids bottlenecks in the NoC with a more regular and predictable mapping of the memory blocks to enable programmers, particularly in the high-performance computing domain, to have full control over coherence traffic. \citet{Horro:Access} developed a simulator for the traffic on the NoC of distributed directory architectures based on the Tejas architectural simulator~\cite{Sarangi:Tejas}, predicting that codes with coherence traffic control would experiment a 20\% decrease in overall traffic over the NoC, yielding more than 50\% latency improvement for the coherence packets. 

In recent years, a number of papers have explored the design of scalable networks-on-chip to support manycore architectures. \citet{Krishna:Scorpio} design a NoC based on an ordered network and a snoopy coherence protocol, and show how congestion increases heavily with the number of cores. \citet{Cuckoo:HPCA11} propose a scalable distributed directory system to alleviate the power and performance problems of sparse and duplicate-tag directories, scaling up to 1,024 cores. \citet{Mishra:NOCS18} identify the importance of the coherence traffic in manycore performance, and show how the memory modes in the Intel KNL can be manipulated to achieve better performance. They neither explore software optimizations to coherence traffic, nor the actual layout of the KNL processor.

Several papers have explored the performance of the KNL architecture, mainly through the analysis of well-known benchmarks, machine learning applications, and parallel workloads~\cite{Azad:KNL,machinelearning:KNL,hadoop:KNL,Bylaska:KNL}. None of these works undertake the analysis of the locality characteristics of the KNL interconnect. \citet{Sabela:IPDPS17} develop a capability model of the cache performance and memory bandwidth of the KNL, characterizing the impact of the different memory and cluster modes. However, this work does not consider the impact of the distributed directory.

Few works focus on data layout optimizations for 2D interconnects. \citet{Bondhugula:PLUTO} propose a polyhedral model and associated optimizations to achieve data locality in these topologies. \citet{Kandemir:DAC15} use a compiler-guided scheme to minimize on-chip network traffic by reducing the distances of cores to data, but without taking into account the effects of a distributed directory.

Finally, other works have proposed ways to discover architectural features, or to automatically tune applications in highly complex systems. \citet{Pingali:Microbenchmarks} developed a set of microbenchmarks to measure parameters of the memory hierarchy. \citet{Kistijantoro:Autotuning} argue that the static discovery of optimal configuration parameters is a fundamentally flawed approach, proposing a configuration interface to specify performance constraints that should be satisfied at runtime. \citet{Hoffman:CALOREE} propose to use automatic learning systems to manage resources towards meeting specific latency and energy constraints.

\section{Conclusion}
\label{sec:Concluding}

Current manycore designs are usually based on replicated IP blocks connected by a high-performance fabric. An example of such an approach is the Intel Mesh interconnect (IM), first featured in the Intel Xeon Phi Knights Landing (KNL) processor~\cite{Sodani:HotChips2015}. The IM is the current interconnection standard in the most advanced Intel processors, including Intel Xeon Scalable servers and the High-End Desktop family of Core-X chips~\cite{Arafa:CascadeLake,Tam:SkyLake}.

In this work, we presented the first complete reverse-engineering of the hardware mapping functions between memory block addresses and the Cache/Home Agent on the KNL, exposing complex bitwise XOR-based functions that can then be exploited at compile-time to further improve data access latency via careful placement. We presented different optimization strategies based on both dynamic and static work scheduling. Extensive experiments quantified the merits and drawbacks of the proposed optimizations, improving memory access latency by leveraging the spatial locality of CHAs. However, our experiments clearly expose the limitations of exploiting such complex XOR-based functions in software, which may ultimately lead to overall performance degradation despite memory latency improvements.

\section*{Acknowledgments}

This research was supported in part by the Ministry of Science and Innovation of Spain (TIN2016-75845-P and PID2019-104184RB-I00, AEI/FEDER/EU, 10.13039/501100011033), the Ministry of Education of Spain (FPU16/00816), and by the U.S. National Science Foundation (award CCF-1750399). CITIC is funded by Xunta de Galicia and FEDER funds of the EU (Centro de Investigación de Galicia accreditation, grant ED431G 2019/01).

\ifCLASSOPTIONcaptionsoff
  \newpage
\fi



\bibliographystyle{plainnat}
{\small\bibliography{TC}}

\begin{thebibliography}{30}
\providecommand{\natexlab}[1]{#1}
\providecommand{\url}[1]{\texttt{#1}}
\expandafter\ifx\csname urlstyle\endcsname\relax
  \providecommand{\doi}[1]{doi: #1}\else
  \providecommand{\doi}{doi: \begingroup \urlstyle{rm}\Url}\fi

\bibitem[Arafa et~al.(2019)Arafa, Fahim, Kottapalli, Kumar, Looi, Mandava,
  Rudoff, Steiner, Valentine, Vedaraman, and Vora]{Arafa:CascadeLake}
M.~Arafa, B.~Fahim, S.~Kottapalli, A.~Kumar, L.P. Looi, S.~Mandava, A.~Rudoff,
  I.M. Steiner, B.~Valentine, G.~Vedaraman, and S.~Vora.
\newblock Cascade lake: Next generation intel xeon scalable processor.
\newblock \emph{IEEE Micro}, 39\penalty0 (2):\penalty0 29--36, 2019.

\bibitem[Augustine et~al.(2019)Augustine, Sarma, Pouchet, and
  Rodríguez]{Gabriel:PLDI19}
T.~Augustine, J.~Sarma, L.-N. Pouchet, and G.~Rodríguez.
\newblock Generating piecewise-regular code from irregular structures.
\newblock In \emph{Proceedings of the 40th ACM SIGPLAN Conference on
  Programming Language Design and Implementation, PLDI}, pages 625--639, 2019.

\bibitem[Azad and Buluç(2017)]{Azad:KNL}
A.~Azad and A.~Buluç.
\newblock A work-efficient parallel sparse matrix-sparse vector multiplication
  algorithm.
\newblock In \emph{Proceedings of the IEEE International Parallel and
  Distributed Processing Symposium, IPDPS}, pages 688--697, 2017.

\bibitem[Byun et~al.(2017)Byun, Kepner, Arcand, Bestor, Bergeron, Gadepally,
  Houle, Hubbell, Jones, Klein, Michaleas, Milechin, Mullen, Prout, Rosa,
  Samsi, Yee, and Reuther]{machinelearning:KNL}
C.~Byun, J.~Kepner, W.~Arcand, D.~Bestor, B.~Bergeron, V.~Gadepally, M.~Houle,
  M.~Hubbell, M.~Jones, A.~Klein, P.~Michaleas, L.~Milechin, J.~Mullen,
  A.~Prout, A.~Rosa, S.~Samsi, C.~Yee, and A.~Reuther.
\newblock Benchmarking data analysis and machine learning applications on the
  {Intel KNL} many-core processor.
\newblock In \emph{Proceedings of the IEEE High Performance Extreme Computing
  Conference, HPEC}, pages 1--6, 2017.

\bibitem[Charles et~al.(2018)Charles, Patil, Ogras, and Mishra]{Mishra:NOCS18}
S.~Charles, C.A. Patil, U.Y. Ogras, and P.~Mishra.
\newblock Exploration of memory and cluster modes in directory-based many-core
  {CMPs}.
\newblock In \emph{Proceedings of the 12th IEEE/ACM International Symposium on
  Networks-on-Chip, NOCS}, pages 1--8, 2018.

\bibitem[Chen et~al.(2017)]{hadoop:KNL}
L.~Chen et~al.
\newblock Benchmarking {Harp-DAAL}: High performance {Hadoop} on {KNL}
  clusters.
\newblock In \emph{Proceedings of the IEEE 10th International Conference on
  Cloud Computing, CLOUD}, pages 82--89, 2017.

\bibitem[Davis and Hu(2011)]{UF-Collection}
T.~A. Davis and Y.~Hu.
\newblock The university of florida sparse matrix collection.
\newblock \emph{ACM Transactions on Mathematical Software}, 38:\penalty0 1--25,
  2011.

\bibitem[Daya et~al.(2014)]{Krishna:Scorpio}
B.K. Daya et~al.
\newblock {SCORPIO}: A 36-core research chip demonstrating snoopy coherence on
  a scalable mesh {NoC} with in-network ordering.
\newblock In \emph{Proceedings of the ACM/IEEE 41st International Symposium on
  Computer Architecture, ISCA}, pages 25--36, 2014.

\bibitem[Farshin et~al.(2019)Farshin, Roozbeh, Maguire, and
  Kostić]{Farshin:RevEng}
A.~Farshin, A.~Roozbeh, G.Q. Maguire, and D.~Kostić.
\newblock {Make the Most out of Last Level Cache in Intel Processors}.
\newblock In \emph{Proceedings of the 14th EuroSys Conference}, pages 8:1--17,
  2019.

\bibitem[Ferdman et~al.(2011)Ferdman, Lotfi-Kamran, Balet, and
  Falsafi]{Cuckoo:HPCA11}
M.~Ferdman, P.~Lotfi-Kamran, K.~Balet, and B.~Falsafi.
\newblock Cuckoo directory: A scalable directory for many-core systems.
\newblock In \emph{Proceedings of the 17th International Conference on
  High-Performance Computer Architecture, HPCA}, pages 169--180, 2011.

\bibitem[Goodman and Hum(2009)]{Goodman:MESIF}
J.R. Goodman and H.H.J. Hum.
\newblock {MESIF}: A two-hop cache coherency protocol for point-to-point
  interconnects.
\newblock Technical report, University of Auckland, 2009.

\bibitem[Horro et~al.(2019{\natexlab{a}})Horro, Kandemir, Pouchet, Rodríguez,
  and Touriño]{Marcos:DAC19}
M.~Horro, M.T. Kandemir, L.-N. Pouchet, G.~Rodríguez, and J.~Touriño.
\newblock Effect of distributed directories in mesh interconnects.
\newblock In \emph{Proceedings of the 56th Annual Design Automation Conference,
  DAC}, pages 51:1--6, 2019{\natexlab{a}}.

\bibitem[Horro et~al.(2019{\natexlab{b}})Horro, Rodríguez, and
  Touriño]{Horro:Access}
M.~Horro, G.~Rodríguez, and J.~Touriño.
\newblock Simulating the network activity of modern manycores.
\newblock \emph{IEEE Access}, 7:\penalty0 81195--81210, 2019{\natexlab{b}}.

\bibitem[Irazoqui et~al.(2015)Irazoqui, Eisenbarth, and Sunar]{Irazoqui:RevEng}
G.~Irazoqui, T.~Eisenbarth, and B.~Sunar.
\newblock {Systematic Reverse Engineering of Cache Slice Selection in Intel
  Processors}.
\newblock In \emph{Euromicro Conference on Digital System Design, DSD}, pages
  629--636, 2015.

\bibitem[Jacquelin et~al.(2017)Jacquelin, Jong, and Bylaska]{Bylaska:KNL}
M.~Jacquelin, W.~De Jong, and E.~Bylaska.
\newblock Towards highly scalable {A}b {I}nitio {M}olecular {D}ynamics ({AIMD})
  simulations on the {Intel Knights Landing} manycore processor.
\newblock In \emph{Proceedings of the IEEE International Parallel and
  Distributed Processing Symposium, IPDPS}, pages 234--243, 2017.

\bibitem[Jeffers et~al.(2016)Jeffers, Reinders, and Sodani]{Sodani:KNL}
J.~Jeffers, J.~Reinders, and A.~Sodani.
\newblock \emph{Intel Xeon Phi Processor High Performance Programming: Knights
  Landing Edition}.
\newblock Morgan-Kauffman, 2016.

\bibitem[Krawczyk(1994)]{LFSR:Krawczyk}
Hugo Krawczyk.
\newblock Lfsr-based hashing and authentication.
\newblock In Yvo~G. Desmedt, editor, \emph{Advances in Cryptology --- CRYPTO
  '94}, pages 129--139, Berlin, Heidelberg, 1994. Springer Berlin Heidelberg.
\newblock ISBN 978-3-540-48658-9.

\bibitem[Liu et~al.(2015)Liu, Kotra, Ding, and Kandemir]{Kandemir:DAC15}
J.~Liu, J.~Kotra, W.~Ding, and M.~Kandemir.
\newblock Network footprint reduction through data access and computation
  placement in {NoC}-based manycores.
\newblock In \emph{Proceedings of the 55th Annual Design Automation Conference,
  DAC}, pages 181:1--6, 2015.

\bibitem[Lu et~al.(2009)Lu, Alias, Bondhugula, Henretty, Krishnamoorthy,
  Ramanujam, Rountev, Sadayappan, Chen, Lin, and Ngai]{Bondhugula:PLUTO}
Q.~Lu, C.~Alias, U.~Bondhugula, T.~Henretty, S.~Krishnamoorthy, J.~Ramanujam,
  A.~Rountev, P.~Sadayappan, Y.~Chen, H.~Lin, and T.-F. Ngai.
\newblock Data layout transformations for enhancing data locality on {NUCA}
  chip multiprocessors.
\newblock In \emph{Proceedings of 18th International Conference on Parallel
  Architectures and Compilation Techniques, PACT}, pages 348--357, 2009.

\bibitem[Maurice et~al.(2015)Maurice, Scouarnec, Neumann, Heen, and
  Francillon]{Maurice:RevEng}
C.~Maurice, N.~Le Scouarnec, C.~Neumann, O.~Heen, and A.~Francillon.
\newblock Reverse engineering intel last-level cache complex addressing using
  performance counters.
\newblock In \emph{Proceedings of the 18th International Symposium on Research
  in Attacks, Intrusions, and Defenses, RAID}, pages 48--65, 2015.

\bibitem[Mishra et~al.(2018)Mishra, Imes, Lafferty, and
  Hoffman]{Hoffman:CALOREE}
N.~Mishra, C.~Imes, J.D. Lafferty, and H.~Hoffman.
\newblock {CALOREE}: Learning control for predictable latency and low energy.
\newblock In \emph{Proceedings of the 23rd International Conference on
  Architectural Support for Programming Languages and Operating Systems,
  ASPLOS}, pages 184--198, 2018.

\bibitem[O'Leary et~al.()O'Leary, Gazizov, Shinsel, Belenov, Matveev, and
  Petunin]{Intel:Advisor}
K.~O'Leary, I.~Gazizov, A.~Shinsel, R.~Belenov, Z.~Matveev, and D.~Petunin.
\newblock {Intel Advisor Roofline Analysis, 2017}.
\newblock
  \url{https://www.codeproject.com/articles/1169323/intel-advisor-roofline-analysis}.
\newblock Accessed: 2020-07-28.

\bibitem[Ramos and Hoefler(2017)]{Sabela:IPDPS17}
S.~Ramos and T.~Hoefler.
\newblock Capability models for manycore memory systems: A case-study with
  {Xeon Phi KNL}.
\newblock In \emph{Proceedings of the IEEE International Parallel and
  Distributed Processing Symposium, IPDPS}, pages 297--306, 2017.

\bibitem[Rodríguez et~al.(2019)Rodríguez, Kandemir, and
  Touriño]{Gabriel:TC19}
G.~Rodríguez, M.T. Kandemir, and J.~Touriño.
\newblock Affine modeling of program traces.
\newblock \emph{IEEE Transactions on Computers}, 68\penalty0 (2):\penalty0
  294--300, 2019.

\bibitem[Sarangi et~al.(2015)Sarangi, Kalayappan, Kallurkar, Goel, and
  Peter]{Sarangi:Tejas}
Smruti~R. Sarangi, Rajshekar Kalayappan, Prathmesh Kallurkar, Seep Goel, and
  Eldhose Peter.
\newblock {Tejas: A Java based versatile micro-architectural simulator}.
\newblock In \emph{International Workshop on Power And Timing Modeling,
  Optimization and Simulation, PATMOS}, 2015.

\bibitem[Sodani(2015)]{Sodani:HotChips2015}
A.~Sodani.
\newblock {Knights Landing (KNL)}: 2nd generation {Intel Xeon Phi} processor.
\newblock In \emph{Proceedings of the IEEE Hot Chips 27 Symposium, HCS}, pages
  1--24, 2015.

\bibitem[Tam et~al.(2018)Tam, Muljono, Huang, Iyer, Royneogi, Satti, Qureshi,
  Chen, Wang, Hsieh, Vora, and Wang]{Tam:SkyLake}
S.M. Tam, H.~Muljono, M.~Huang, S.~Iyer, K.~Royneogi, N.~Satti, R.~Qureshi,
  W.~Chen, T.~Wang, H.~Hsieh, S.~Vora, and E.~Wang.
\newblock {SkyLake-SP}: A 14nm 28-core {Xeon} processor.
\newblock In \emph{Proceedings of the IEEE International Solid-State Circuits
  Conference, ISSCC}, pages 34--36, 2018.

\bibitem[Viswanathan(2014)]{Intel:HWPF}
K.~Viswanathan.
\newblock {Disclosure of Hardware Prefetcher Control on Some Intel®
  Processors}.
\newblock
  \url{https://software.intel.com/content/www/us/en/develop/articles/disclosure-of-hw-prefetcher-control-on-some-intel-processors.html},
  2014.
\newblock Last accessed: August 2020.

\bibitem[Wang et~al.(2018)Wang, Li, Hoffman, Lu, Sentosa, and
  Kistijantoro]{Kistijantoro:Autotuning}
S.~Wang, C.~Li, H.~Hoffman, S.~Lu, W.~Sentosa, and A.I. Kistijantoro.
\newblock Understanding and auto-adjusting performance-sensitive
  configurations.
\newblock In \emph{Proceedings of the 23rd International Conference on
  Architectural Support for Programming Languages and Operating Systems,
  ASPLOS}, pages 154--168, 2018.

\bibitem[Yotov et~al.(2005)Yotov, Pingali, and
  Stodghill]{Pingali:Microbenchmarks}
K.~Yotov, K.~Pingali, and P.~Stodghill.
\newblock Automatic measurement of memory hierarchy parameters.
\newblock In \emph{Proceedings of the International Conference on Measurement
  and Modeling of Computer Systems, SIGMETRICS}, pages 181--192, 2005.

\end{thebibliography}

%
%

%

\begin{IEEEbiography}[{\includegraphics[width=1in,height=1.25in,clip,keepaspectratio]{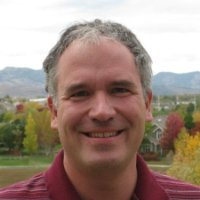}}]{Steve Kommrusch}
  is a PhD student in the Department of Computer Science at Colorado State University (CSU). His research is on machine learning techniques for syntactic understanding and generation of computer code, AI safety, and verifiable ways to use machine learning. Prior to beginning his computer science PhD work, he worked in the field of computer hardware design attaining the level of Fellow design engineer at Advanced Micro Devices. He has 31 patents granted in the fields of computer graphics, system-on-chip design, silicon debugging, low-power processors, and asynchronous clocking techniques.

\end{IEEEbiography}

\begin{IEEEbiography}[{\includegraphics[width=1in,height=1.25in,clip,keepaspectratio]{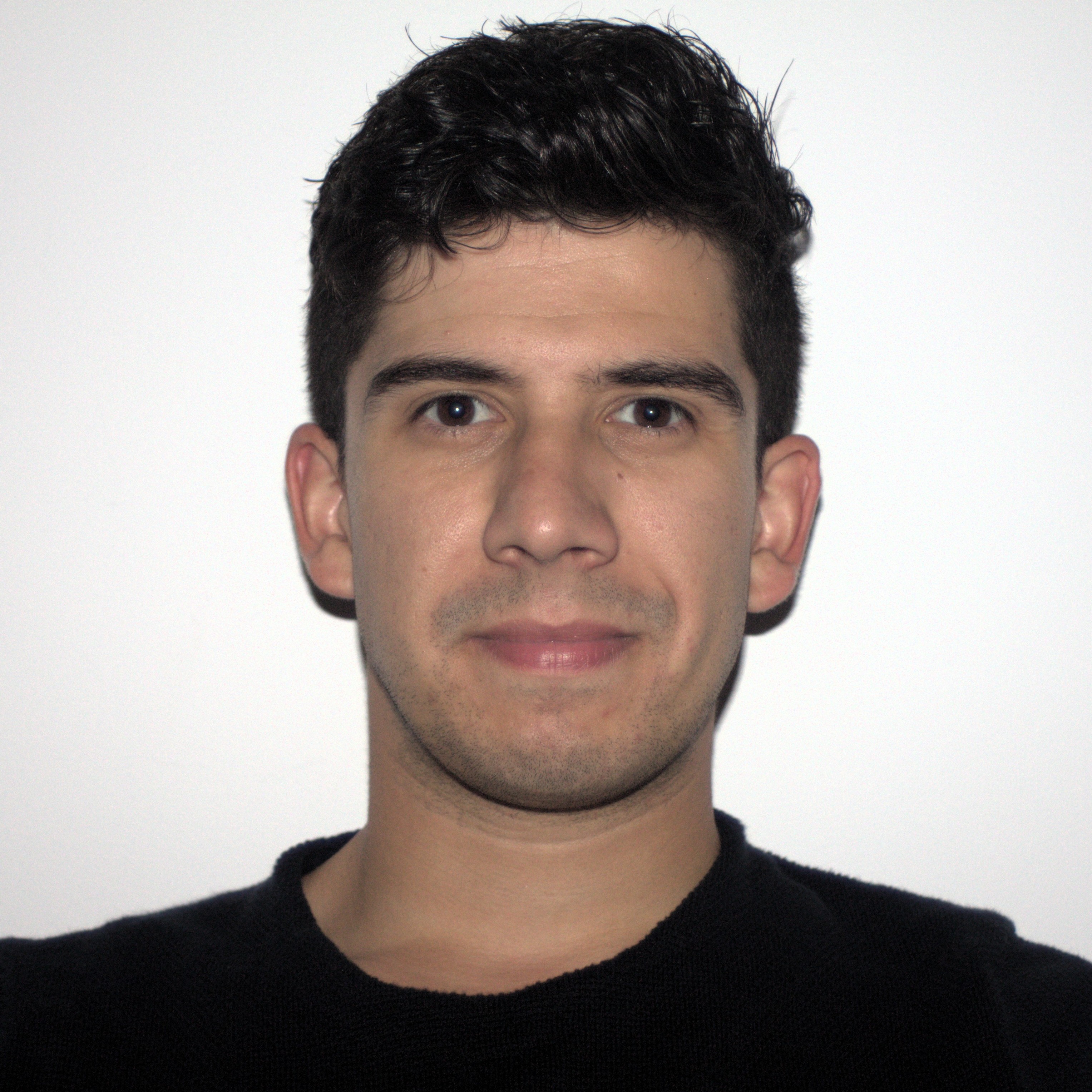}}]{Marcos Horro}
	received the B.S. and M.S. degrees in computer science from the 
	Universidade da Coruña, Spain, in 2016 and 2018, respectively. He is 
	currently a PhD candidate with the Department of Computer Engineering, 
	Universidade da Coruña. His main research interests are in the area of HPC, 
	computer architecture focused on the performance evaluation and 
	optimization of heterogeneous memory systems and compilers. His homepage is 
	\url{http://gac.udc.es/~horro}.
\end{IEEEbiography}

\begin{IEEEbiography}[{\includegraphics[width=1in,height=1.25in,clip,keepaspectratio]{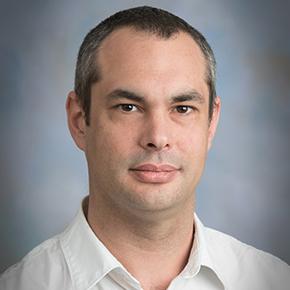}}]{Louis-No{\"e}l Pouchet}
  is an Associate Professor of Computer Science at Colorado State University, Fort Collins, CO, USA, with a joint appointment in the Electrical and Computer Engineering Department. He is working on pattern-specific languages and compilers for scientific computing, and has designed numerous approaches using optimizing compilation to effectively map applications to CPUs, GPUs, FPGAs, and System-on-Chips. His work spans a variety of domains including compiler optimization design especially in the polyhedral compilation framework, high-level synthesis for FPGAs and SoCs, and distributed computing.
  He is the author of the PolyOpt and PoCC compilers, and of the PolyBench benchmarking suite.
\end{IEEEbiography}

\begin{IEEEbiography}[{\includegraphics[width=1in,height=1.25in,clip,keepaspectratio]{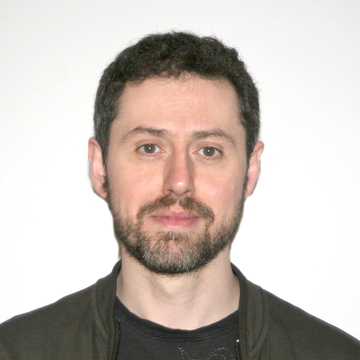}}]{Gabriel Rodríguez}
  is an Associate Professor in the Department of Computer Engineering at the University of A Coruña, where he is a member of the Computer Architecture Group and CITIC Research. His main research interests are in the field of optimizing compilers, architectural support for high performance computing, and power-aware computing. His homepage is \url{http://gac.udc.es/~gabriel}.
  
\end{IEEEbiography}

%
\begin{IEEEbiography}[{\includegraphics[width=1in,height=1.25in,clip,keepaspectratio]{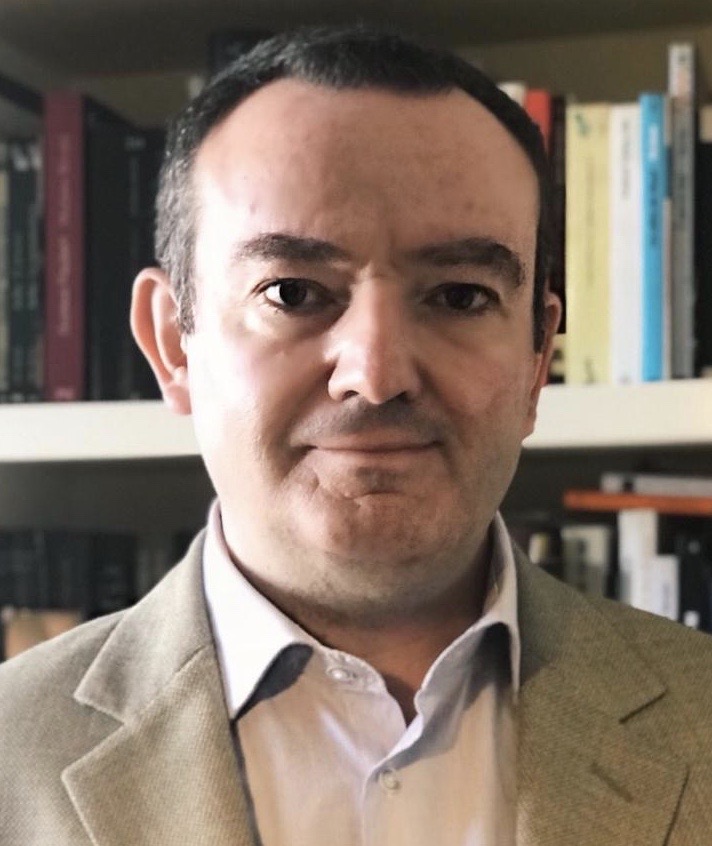}}]{Juan Touriño}
	(M'01-SM'06) is a Full Professor in the Department of Computer Engineering at the University of A Coruña, where he leads the Computer Architecture Group. He has extensively published in the area of High Performance Computing (HPC): programming languages and compilers for HPC, high performance architectures and networks, parallel algorithms and applications, etc. He is coauthor of more than 170 papers on these topics in international conferences and journals. He has also served in the Program Committee of 70 international conferences. His homepage is \url{http://gac.udc.es/\~juan}.
\end{IEEEbiography}
\vfill

\end{document}